\documentclass[trackchanges, twocolumn]{aastex701}

\usepackage{verbatim}
\usepackage{booktabs}
\usepackage{mhchem}
\usepackage{textcomp}

\begin{document}

\title{exoALMA XXIII. Estimating Disk and Planet Properties from Dust Morphologies with DBNets2.0}

\author[orcid=0009-0007-6448-658X,sname='Alessandro Ruzza']{Alessandro Ruzza}
\affiliation{Dipartimento di Fisica, Universit\`a degli Studi di Milano, Via Celoria 16, 20133 Milano, Italy}
\email[]{alessandro.ruzza@unimi.it}  

\author[orcid=0000-0002-2357-7692,sname='Giuseppe Lodato']{Giuseppe Lodato}
\affiliation{Dipartimento di Fisica, Universit\`a degli Studi di Milano, Via Celoria 16, 20133 Milano, Italy}
\email[]{giuseppe.lodato@unimi.it} 

\author[orcid=0000-0003-4853-5736 ,sname='Giovanni Pietro Rosotti']{Giovanni Rosotti}
\affiliation{Dipartimento di Fisica, Universit\`a degli Studi di Milano, Via Celoria 16, 20133 Milano, Italy}
\email[]{giovanni.rosotti@unimi.it} 

\author[orcid=0000-0001-5032-1396, sname='Philip J. Armitage']{Philip Armitage}
\affiliation{Center for Computational Astrophysics, Flatiron Institute, 162 Fifth Ave, New York, NY 10010, USA}
\affiliation{Department of Physics and Astronomy,
Stony Brook University,
Stony Brook, NY 11794, USA
}
\email[]{parmitage@flatironinstitute.org} 

\author[orcid=0000-0003-4689-2684, sname='Stefano Facchini']{Stefano Facchini}
\affiliation{Dipartimento di Fisica, Universit\`a degli Studi di Milano, Via Celoria 16, 20133 Milano, Italy}
\email[]{stefano.facchini@unimi.it}

\author[0000-0003-2253-2270]{Sean M. Andrews}
\affiliation{Center for Astrophysics, Harvard \& Smithsonian, Cambridge, MA 02138, USA}
\email[]{sandrews@cfa.harvard.edu}

\author[0000-0001-7258-770X]{Jaehan Bae}
\affiliation{Department of Astronomy, University of Florida, Gainesville, FL 32611, USA}
\email[]{jbae@ufl.edu}

\author[0000-0001-6378-7873]{Marcelo Barraza-Alfaro}
\affiliation{Department of Earth, Atmospheric, and Planetary Sciences, Massachusetts Institute of Technology, Cambridge, MA 02139, USA}
\email[]{mbarraza@mit.edu}

\author[0000-0000-0000-0000]{Myriam Benisty}
\affiliation{Universit\'e C\^ote d'Azur, Observatoire de la C\^ote d'Azur, CNRS, Laboratoire Lagrange, 06304 Nice, France}
\affiliation{Max-Planck Institute for Astronomy (MPIA), Königstuhl 17, 69117 Heidelberg, Germany}
\email[]{benisty@mpia.de}

\author[0000-0003-2045-2154]{Pietro Curone} 
\affiliation{Departamento de Astronom\'ia, Universidad de Chile, Camino El Observatorio 1515, Las Condes, Santiago, Chile}
\email[]{pcurone@das.uchile.cl}

\author[0000-0003-4679-4072]{Daniele Fasano} 
\affiliation{Universit\'e C\^ote d'Azur, Observatoire de la C\^ote d'Azur, CNRS, Laboratoire Lagrange, 06304 Nice, France}
\affiliation{Max-Planck Institute for Astronomy (MPIA), Königstuhl 17, 69117 Heidelberg, Germany}
\email[]{dafasano@mpia.de}

\author[0000-0002-8138-0425]{Cassandra Hall} 
\affiliation{Department of Physics and Astronomy, The University of Georgia, Athens, GA 30602, USA}
\affiliation{Center for Simulational Physics, The University of Georgia, Athens, GA 30602, USA}
\affiliation{Institute for Artificial Intelligence, The University of Georgia, Athens, GA, 30602, USA}
\email[]{cassandra.hall@uga.edu}

\author[0000-0001-7641-5235]{Thomas Hilder} 
\affiliation{School of Physics and Astronomy, Monash University, Clayton VIC 3800, Australia}
\email[]{Thomas.Hilder@monash.edu}

\author[0000-0001-8446-3026]{Andr\'es F. Izquierdo} 
\affiliation{Department of Astronomy, University of Florida, Gainesville, FL 32611, USA}
\affiliation{NASA Hubble Fellowship Program Sagan Fellow}
\email[]{andres.izquierdo.c@gmail.com}

\author[0000-0003-4663-0318]{Cristiano Longarini} 
\affiliation{Institute of Astronomy, University of Cambridge, Madingley Road, CB3 0HA, Cambridge, UK}
\email[]{cl2000@cam.ac.uk}

\author[0000-0002-1637-7393]{François. Ménard} 
\affiliation{Univ. Grenoble Alpes, CNRS, IPAG, B38000 Grenoble, France}
\email[]{francois.menard@univ-grenoble-alpes.fr}

\author[0000-0001-5907-5179]{Christophe Pinte}
\affiliation{Univ. Grenoble Alpes, CNRS, IPAG, B38000 Grenoble, France}
\affiliation{School of Physics and Astronomy, Monash University, Clayton VIC 3800, Australia}
\email[]{christophe.pinte@gmail.com}

\author[0000-0002-0491-143X]{Jochen Stadler} 
\affiliation{Universit\'e C\^ote d'Azur, Observatoire de la C\^ote d'Azur, CNRS, Laboratoire Lagrange, 06304 Nice, France}
\affiliation{European Southern Observatory, Karl-Schwarzschild-Str. 2, D-85748 Garching bei M\"unchen, Germany}
\email[]{jochen.stadler@oca.eu}

\author[0000-0003-1534-5186]{Richard Teague}
\affiliation{Department of Earth, Atmospheric, and Planetary Sciences, Massachusetts Institute of Technology, Cambridge, MA 02139, USA}
\email[]{rteague@mit.edu}

\author[0000-0002-8590-7271]{Jason Terry}
\affiliation{University of Oxford: Oxford, Oxfordshire, GB}
\email[]{jpterry@uga.edu}


\author[0000-0003-1526-7587	]{David J. Wilner} 
\affiliation{Center for Astrophysics | Harvard \& Smithsonian, Cambridge, MA 02138, USA}
\email[]{dwilner@cfa.harvard.edu}

\author[0000-0002-7501-9801]{Andrew J. Winter}
\affiliation{Astronomy Unit, School of Physics and Astronomy, Queen Mary University of London, London E1 4NS, UK}
\email[]{andrew.winter@oca.eu}

\author[0000-0001-8002-8473	]{Tomohiro C. Yoshida} 
\affiliation{National Astronomical Observatory of Japan, 2-21-1 Osawa, Mitaka, Tokyo 181-8588, Japan}
\email[]{tomohiroyoshida.astro@gmail.com}

\author[0000-0001-9319-1296	]{Brianna Zawadzki} 
\affiliation{Department of Astronomy, Van Vleck Observatory, Wesleyan University, 96 Foss Hill Drive, Middletown, CT 06459, USA}
\affiliation{Department of Astronomy \& Astrophysics, 525 Davey Laboratory, The Pennsylvania State University, University Park, PA 16802, USA}
\email[]{bzawadzki@wesleyan.edu}

\defcitealias{Ruzza2025DBNets2.0:Discs}{R25}

\begin{abstract}

The exoALMA large program provided an unprecedented view of the morphology and kinematics of 15 circumstellar disks, offering a biased but homogenous and well-characterized sample for population-level analysis. Continuum observations revealed numerous dust substructures, known to be potential signatures of embedded planets. We analyze the observed dust morphologies with the simulation-based inference tool DBNets2.0, assuming these are due to embedded planets at fixed locations, to infer the system properties.
We estimate the putative planet mass, the disk $\alpha$-viscosity, scale-height, and dust Stokes number that would reproduce 19 substructures in 13 of the 15 exoALMA disks. We compare our results with literature estimates derived with different methods, and find good agreement in most cases. We further explore the implications of the inferred disk properties for accretion, showing that for the Herbig stars in our sample, the implied viscous accretion timescales are too long to account for their observed stellar accretion rates. Regarding planet migration, our results favor inward migration, with only three putative planets expected to migrate outward. Finally, we check for correlations of the inferred disk and planet properties with the disks' gas-to-dust mass ratio, non-axisymmetry index, and masses of the gas, dust, and host stars, finding no remarkable trend.

\end{abstract}


\footnote{Corresponding author: \href{mailto:alessandro.ruzza@unimi.it}{alessandro.ruzza@unimi.it}}


\section{Introduction} 

Substructures, such as gaps, rings, and cavities, are ubiquitous in protoplanetary and transition disk continuum observations (e.g. \citealt{ALMAPartnership2015TheRegion,Isella2016RingedALMA, Clarke2018High-resolutionAu, Dipierro2018RingsALMA, Andrews2018TheOverview, vanTerwisga2018V1094Star, Huang2020ADisk, Bae2023StructuredDisks, Curone2025ExoALMA.Emission}). Although several mechanisms have been proposed to explain their origin (e.g. \citealt{Hawley2001GlobalDisks, Barge2017GapsDust, Dullemond2018Dust-drivenDisks, Hu2019NonidealRings, Bae2023StructuredDisks}), planet-disk interaction remains one of the most promising and discussed (e.g. \citealt{Dipierro2015OnTau, Rosotti2016TheObservations, Zhang2018TheInterpretation, Lodato2019TheDiscs, Ruzza2024DBNets:Discs, Ruzza2025DBNets2.0:Discs}) observationally confirmed in the two cases of PDS~70 \citep[e.g.,][]{Keppler2019HighlyALMA, Isella2019DetectionProtoplanets} and WISPIT~2 \citep{Capelleveen2025WIde2}.
In this scenario, the observed dust substructures emerge due to the gravitational interaction with embedded young planets, resulting in the formation of gaps and rings whose morphological features depend on the physical properties of the system \citep{Rosotti2016TheObservations, Kanagawa2016MassWidth, Kanagawa2017ModellingDisks, Dipierro2017AnDiscs, Lodato2019TheDiscs}.
Substructures can thus serve as an indirect probe for disk and planet properties that are still poorly constrained.

The large number of available observations enables gathering samples suitable for population-level statistically significant analysis and discussion, as done for instance in \cite{Lodato2019TheDiscs} and  \citet{Ruzza2024DBNets:Discs, Ruzza2025DBNets2.0:Discs}.
 The estimation is done with simulation-based inference (SBI), which consists of fitting data with a model that is either the direct output of numerical simulations or a representation of that output. This process has been conducted manually by comparing synthetic observations to actual data \citep{Dipierro2018RingsALMA, Fedele2018ALMA209, Toci2019Long-lived169142, Veronesi2020IsPlanetb, Zhang2018TheInterpretation} or through the calibration of empirical relationships \citep{Kanagawa2016MassWidth, Lodato2019TheDiscs, Dong2017WhatPlanet}. A more recent alternative has been the use of machine-learning techniques \citep{Auddy2020ADisks, Auddy2021DPNNet-2.0.Gaps, Auddy2022UsingDisks, Zhang2022PGNets:Discs, Ruzza2024DBNets:Discs, Mao2024Disk2Planet:Systems, Ruzza2025DBNets2.0:Discs, ShafaatMahmud2025InferringAutoencoder}, which has shown significant advantages over the other methods, combining the same speed and ease of use of empirical relations with the higher expressiveness of neural networks and the possibility, as done when fine-tuning simulations, of exploiting the entire observed morphology. These methods have demonstrated high accuracy on synthetic datasets and, in most cases, provided reliable quantification of the uncertainties.
 
The exoALMA Large Program \citep{Teague2025ExoALMA.Products} results represent a rare and valuable case of a homogenous sample of disks' dust continuum observations. All of them present substructures in the millimeter continuum identified and characterized by \cite{Curone2025ExoALMA.Emission}, with the sole exception of PDS~66. It should be noted, however, that substructures were also observed in this disk at higher resolution in similar wavelengths \citep[ALMA band 7,][]{Aguayo2025ConfirmationObservations} and at longer \citep[ALMA band 3][]{Ribas2025ADisk}, and shorter \citep{Wolff2016THEIMAGER} wavelengths, which trace different populations of dust grains.
In this paper, we aim to investigate what population of planets would be consistent with these substructures, assuming that each is the result of the disk interaction with an embedded young planet. We estimate, for all the objects in this sample, both disks and planets' physical properties. We then compare them with literature constraints and analyze their implications for both the observability and evolution of these systems.

To this end, we use DBNets2.0 \citep[hereafter \citetalias{Ruzza2025DBNets2.0:Discs}]{Ruzza2025DBNets2.0:Discs}, a state-of-the-art SBI method that exploits machine learning techniques to enable a quick inference of the planet and disk properties that have a role in determining the observed morphology. Using neural posterior estimation \citep[NPE; see, e.g.,][]{Cranmer2020TheInference}, the full joint posterior for these properties is inferred, revealing degeneracies and enabling uncertainty quantification. 

Further details regarding DBNets2.0 are provided in Sect. \ref{sec:dbnets20}. Section \ref{sec:sample} presents our motivation for the selection of this sample and our strategy for the analysis of each source. Results and systematic comparisons with previous studies are presented in Sect. \ref{sec:results} with further discussion in Sect. \ref{sec:discussion}. In Sect. \ref{sec:conclusion} we draw our conclusions.

\section{DBNets2.0}
\label{sec:dbnets20}

DBNets2.0 is a simulation-based inference pipeline for rapidly fitting dust substructures, such as gaps, rings and cavities, observed in disks' dust continuum emission. Assuming the planet-disk interaction scenario, the tool takes as input the dust continuum observation, its angular resolution, and the assumed planet location, and infers the mass of the putative planet responsible for the observed substructures. In addition, it constrains three disk properties that can influence their morphology: the disk $\alpha$-viscosity \citep{Shakura1973BlackAppearances}, gas aspect ratio ($h=H/R$), and dust Stokes number (St). To do that, DBNets2.0 uses convolutional neural networks (CNNs) and normalizing flows (NF, see, e.g., \citealt{Papamakarios2017MaskedEstimation}), which allow the inference of the full posterior distribution for the target properties, enabling an uncertainty-aware analysis and unveiling all existing degeneracies. The employed methods make the tool fast to apply to new data and easily scalable to larger datasets. Additionally, the extensive testing carried out in \citetalias{Ruzza2025DBNets2.0:Discs} on synthetic data demonstrated the accuracy of the inference results.

This pipeline is ultimately fitting data with a specific model, and all the inferred posteriors are to be understood as conditional upon it. This is not unique of SBI methods as the same is generally true for any technique of Bayesian inference with a model. In SBI, the model is implicitly defined by the training dataset, which, in this case, is composed of disks synthetic observations generated from the results of 2D hydrodynamical simulations. To clarify the assumptions underlying our results, we report in Appendix \ref{app:dbnetsmodel} the key steps and simplifications used to generate these data. Further details are provided in \cite{Ruzza2024DBNets:Discs, Ruzza2025DBNets2.0:Discs}.
The inferred distributions cannot account for scenarios that are not included in this model. Therefore, it must be assessed if, or in which conditions, this model is appropriate to analyze the data.
When the outlined assumptions can be made safely is, in general, an open question that goes beyond the scope of this work. We try to touch upon this issue first, more generally, in Appendix \ref{app:dbnetsmodel} and then in more detail by discussing, in Appendix \ref{app:allsources}, each disk separately in light of all the information available from previous studies. 

To support the analysis of whether the underlying model can reproduce the data, DBNets2.0 provides a metric called “confidence score" (CS) that quantifies, with a value between 0 and 1, the similarity between the input continuum observation and the synthetic data within DBNets2.0 training sample (the exact definition can be found in Appendix \ref{app:dbnetsmodel}).
\citetalias{Ruzza2025DBNets2.0:Discs} identified 0.6 as a loose threshold between input data from the same distribution of the training dataset and random or out-of-distribution data. It was advised to treat with additional caution (or reject), any result with a CS less than this threshold. However, a small overlap around this threshold between in-distribution and out-of-distribution data was shown. Hence, in this work, since lower values are still close to this threshold, we adopt a conservative approach, considering estimates with CS~$< 0.6$ as uncertain. We exclude these points from the presentation and discussion of aggregated results, but we still present them in plots where they can be clearly marked.

Additionally, because DBNets2.0, as any ML method, cannot safely extrapolate beyond the parameter space where it was trained, posteriors that extend beyond it should be regarded as non-constraining. We identify as such all
estimates whose 2.5 and 97.5 percentiles (corresponding
to $2\sigma$ for a Gaussian) fall outside of the respective prior support. As for estimates with CS$<0.6$, unconstrained estimates are
removed from our presentation and discussion of aggregated results, while they are marked as uncertain in
plots that distinguish individual estimates.

\section{Sample selection and methodology}
\label{sec:sample}

For this study, we considered all fiducial Band 7 continuum observations of the exoALMA Large Program, which were CLEANed, calibrated, and made publicly available by the exoALMA collaboration \citep{Teague2025ExoALMA.Products, Loomis2025ExoALMA.Pipeline, Curone2025ExoALMA.Emission}.
It is important to note that the exoALMA sample is inherently affected by selection biases, leading to an overrepresentation of brighter and more extended disks \citep[for details see][]{Teague2025ExoALMA.Products}.

DBNets2.0 requires as input a proposed planet location to fit the observed substructures. 
In this paper, we separately consider each disk and propose the locations of embedded planets based on all available information, including dust morphology and relevant literature. Table \ref{tab:allres} presents a summary of the proposed planet locations derived from this discussion, while Appendix \ref{app:allsources} outlines the considerations that guided these choices. 

As a general strategy, we consider as viable planet locations those of gaps and cavities identified by \cite{Curone2025ExoALMA.Emission} through the characterization of the azimuthally symmetric \texttt{frank} \citep{Jennings2020Frankenstein:Process} fits of the disks' continuum emission. We also follow \cite{Curone2025ExoALMA.Emission} characterization when distinguishing between gaps and cavities. We note that both types of substructures are equally represented in the DBNets2.0 training dataset \citep[see][]{Ruzza2024DBNets:Discs, Ruzza2025DBNets2.0:Discs}.
In addition to \cite{Curone2025ExoALMA.Emission} selection criteria, we exclude substructures that are not visible in the azimuthally averaged radial profiles obtained from the CLEANed images, such as the D63 gap in J1842, and consider shallow gaps in low signal-to-noise observations only if a planet at the same location was already proposed in previous studies.

In some disks with multiple gaps, we assume multiple planet locations, interpreting the results as described in Appendix \ref{app:dbnetsmodel}.
In the case of cavities, planets could be located over a wider range of radii, and DBNets2.0 output is sensitive to this choice (see Sect. \ref{sec:rp}). We select the putative planet's location based on previous studies that proposed detections of planet signatures such as direct thermal emission or kinematic perturbations, and exclude from our main discussions the cavity of J1842, which lacks such constraints. However, because these detections remain debated, in Appendix \ref{app:rpdeg} we provide DBNets2.0 estimates as a function of the assumed planet location, including J1842 in this analysis.

\begin{table*}[]
\begin{center}


\begin{tabular}{lllllllll}
    \toprule 
        Disk name & $M_\star$ [$M_\odot$]& $R_p$ [au]& $R_\text{edge}$ [au] & log~$\alpha$ & $h$ & log~St & $M_p$ [M$_J$] & CS \\
    \midrule 
    AA Tau & 0.79 & 11, & & $-3.69^{+0.35}_{-0.39}$, & $-$, & $-$, & $2.22^{+2.17}_{-1.21}$, & 0.64, \\ 
     & & 72& & $-3.63^{+0.28}_{-0.23}$ & $0.07^{+0.00}_{-0.00}$ & $-1.51^{+0.12}_{-0.13}$ & $0.07^{+0.02}_{-0.02}$ & 0.84\\ 
    
    CQ Tau & 1.40 & \underline{20} & 41 & $-3.71^{+0.49}_{-0.59}$ & $0.04^{+0.01}_{-0.01}$ & $-2.49^{+0.40}_{-0.36}$ & $2.34^{+0.93}_{-1.02}$ & 0.60\\ 
    
    DM Tau & 0.45 & 14, & & $-$, & $-$, & $-$,  & $-$,  & 0.66,\\ 
     & & 72& &  $-2.26^{+0.20}_{-0.20}$ &  $0.05^{+0.01}_{-0.01}$ &  $-2.07^{+0.13}_{-0.12}$ & $0.05^{+0.02}_{-0.01}$ & 0.96\\ 
    
*HD 135344B & 1.61 & \underline{28},& 51 & $-2.78^{+0.29}_{-0.38}$, & $0.04^{+0.01}_{-0.01}$, & $-2.64^{+0.27}_{-0.34}$,  & $2.46^{+0.74}_{-0.68}$, & 0.49,\\ 
& & 66 & & $-3.82^{+0.20}_{-0.18}$ & $0.07^{+0.01}_{-0.01}$ &  $-1.56^{+0.13}_{-0.14}$ & $0.23^{+0.07}_{-0.06}$ & 0.68\\

HD 143006 & 1.56 & 22, & &$-$,  & $0.07^{+0.04}_{-0.01}$, & $-2.76^{+0.37}_{-0.41}$, & $2.59^{+1.41}_{-1.36}$,  & 0.73,\\ 
& & 52& &  $-3.40^{+0.33}_{-0.28}$ &  $0.08^{+0.01}_{-0.01}$ &  $-1.28^{+0.22}_{-0.23}$ & $0.33^{+0.14}_{-0.09}$ & 0.63\\ 

HD 34282 & 1.61 & \underline{47}& 124 & $-3.72^{+0.20}_{-0.18}$ & $0.07^{+0.02}_{-0.02}$ & $-1.89^{+0.36}_{-0.35}$ & $7.14^{+3.33}_{-1.76}$ & 0.67\\ 

J1604 & 1.29 & \underline{41} & 82 & $-2.99^{+0.31}_{-0.31}$ & $0.06^{+0.01}_{-0.01}$ & $-2.05^{+0.37}_{-0.36}$ & $7.91^{+1.52}_{-1.37}$ & 0.86\\ 

J1615 & 1.14 & 83 & & $-2.35^{+0.19}_{-0.17}$ & $0.07^{+0.01}_{-0.01}$ & $-2.46^{+0.19}_{-0.20}$ & $0.29^{+0.08}_{-0.07}$ & 0.95\\ 

J1852 & 1.03 & 31& & $-2.38^{+0.22}_{-0.27}$ & $0.07^{+0.01}_{-0.01}$ & $-1.82^{+0.22}_{-0.28}$ & $2.12^{+0.72}_{-0.66}$ & 0.61\\ 

*LkCa 15 & 1.14 & \underline{42},&68 & $-2.93^{+0.17}_{-0.16}$, & $0.08^{+0.01}_{-0.01}$, & $-2.52^{+0.19}_{-0.17}$, & $1.63^{+0.28}_{-0.26}$, & 0.54,\\
 & & 86 &&  $-3.76^{+0.25}_{-0.22}$ & $0.06^{+0.00}_{-0.00}$ &  $-1.45^{+0.12}_{-0.14}$ & $0.07^{+0.02}_{-0.02}$ & 0.72\\

*MWC 758 & 1.40 & 30& & $-3.55^{+0.31}_{-0.37}$ & $0.04^{+0.01}_{-0.01}$ & $-2.23^{+0.30}_{-0.32}$ & $2.35^{+0.57}_{-0.43}$ & 0.52\\ 
SY Cha & 0.77 & 33& & $-3.78^{+0.17}_{-0.17}$ & $0.07^{+0.02}_{-0.01}$ & $-1.81^{+0.34}_{-0.34}$ & $2.47^{+0.82}_{-0.63}$ & 0.61\\ 

V4046 Sgr & 1.73 & 8,& & $-3.75^{+0.34}_{-0.38}$,  & $-$, & $-$, & $3.73^{+3.25}_{-1.96}$, & 0.86,\\ 
& & 20& & $-3.32^{+0.18}_{-0.17}$ & $0.09^{+0.01}_{-0.01}$ &  $-2.83^{+0.17}_{-0.17}$ & $1.49^{+0.36}_{-0.28}$ & 0.67\\ 

    \bottomrule
    \end{tabular}
    \caption{Analyzed disks, putative planet locations, and inferred best estimates for disks and planet properties. Putative planet locations within cavities are underlined. In these cases, we also report the cavity edge ($R_\text{edge}$), which we reference in section \ref{sec:rp}. Confidence scores (CS) quantify, with a value between 0 and 1, the similarity of the input observation to DBNets2.0 training data. Following \citetalias{Ruzza2025DBNets2.0:Discs}, CS$\leq$0.6 indicates poor representation of the respective dust morphology in DBNets2.0 training data. Therefore, they must be treated with additional caution and uncertainty. We mark these cases with an asterisk before the disk name. Because lower CS values are still close to this threshold, we still provide all our results. Unconstrained estimates are not reported in the table.}
    \label{tab:allres}
    \end{center}
\end{table*}

For each disk and putative planet, the output is an estimate of the posterior distribution $p(\alpha, h, \text{St}, M_p/M_\star| x, \mathcal{M})$ conditioned on the observation $x$ and the underlying model $\mathcal{M}$ discussed in section \ref{sec:dbnets20}. All analyses and results presented here are based on $10^6$
 samples drawn from each inferred posterior distribution. The conversion from $M_p/M_\star$ to $M_p$ is done using dynamical stellar masses derived by \cite{Izquierdo2025ExoALMA.Disks}.

\section{Results and comparison with literature}
\label{sec:results}

\begin{figure*}
    \centering
    \includegraphics[width=\linewidth]{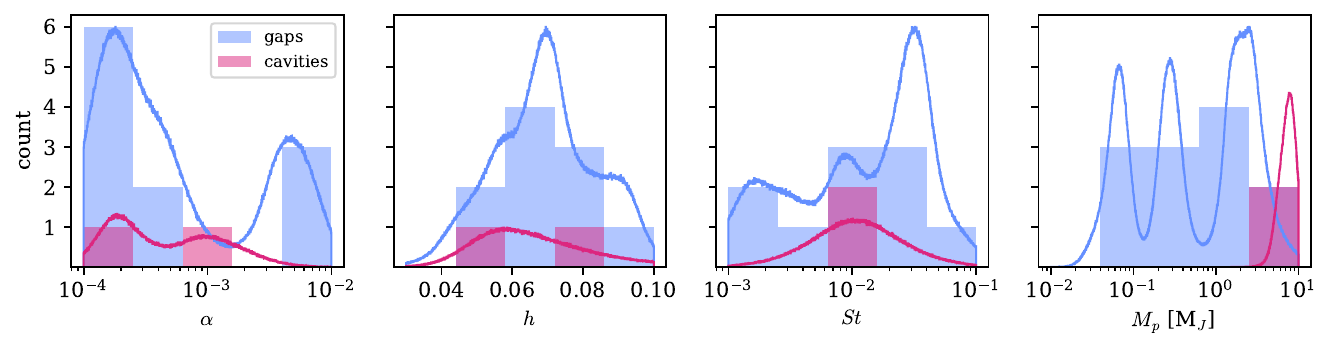}
    \caption{Distributions of the four inferred properties for the full sample, excluding results considered uncertain or unconstrained. The shaded histograms show the medians of the individual posterior distributions, while the overlaid curves represent the stacked posteriors, normalized for comparison. Y-axis ticks refer to the histograms. Contributions from putative planets located in gaps (blue) and in cavities (pink) are shown separately.}
    \label{fig:allsamp_hist}
\end{figure*}

Individual results for each proposed planet are reported in Table \ref{tab:allres} and Appendix \ref{app:all_res}. We also report the ``confidence scores" (CS) of DBNets2.0 estimates which, in three cases, fall below the rejection threshold (CS=0.6) set by \citetalias{Ruzza2025DBNets2.0:Discs}. These are three estimates for putative planets in the inner gap of MWC~758 and in the cavities of HD~135344~B and LkCa~15. Additionally, unconstrained estimates are removed from table \ref{tab:allres} and marked in red in Appendix \ref{app:all_res}. They mainly occur for substructures at small radial distances from the host star. In the rest of the paper, following the approach already mentioned in Sect.~\ref{sec:sample}, unconstrained estimates and those with $\text{CS}<0.6$ are removed from our presentation and discussion of aggregated results, while they are marked as uncertain in plots that distinguish individual estimates.

Figure \ref{fig:allsamp_hist} shows the distributions of the four inferred properties for the full sample, excluding uncertain estimates. No noteworthy trends are evident in the distributions of $h$ and St. The inferred $\alpha$-viscosity distribution presents two peaks at the two ends of DBNets2.0 prior (see Appendix \ref{app:scope}), with an higher number of sources showing evidence of low viscosity, in line with the results found on the larger sample analysed in \citetalias{Ruzza2025DBNets2.0:Discs} and with several literature studies (e.g. \citealt{Flaherty2015WEAKOBSERVATIONS, Zhang2018TheInterpretation, Teague2016MeasuringLimitations, Rosotti2023EmpiricalDiscs}). Among the three systems with high viscosity estimates is the noteworthy case of DM Tau, further discussed in section \ref{sec:dmtau}, where there is evidence of a high level of turbulence   \citep{Guilloteau2012ChemistryDisks, Flaherty2020MeasuringSgr}. We checked but did not observe any common characteristics distinguishing the high and low-viscosity groups that could account for the observed bimodality. 

In contrast to \citetalias{Ruzza2025DBNets2.0:Discs}, where 83\% of planets had masses below $1\text{M}_J$, in this case the distribution of planet masses extends significantly to higher values. This result may reflect the exoALMA target-selection strategy, which favored the most massive disks, arguably more likely to host massive observable companions. The three peaks distribution is due to the combination of the low number and high precision of our mass estimates.
Cavities are found to suggest the presence of more massive companions ($\geq 1$~M$_J$) while no clear distinction is observed for the other disk properties. It should be noted that for cavities, after removal of those unconstrained or uncertain, only two estimates are shown for each property, although they do not all correspond to the same two putative planets. The details of which estimates are shown, with precise values, can be found in Table \ref{tab:allres}.

\subsection{Planet masses}
\label{sec:resump}

\begin{figure*}
    \centering
    \includegraphics[width=\linewidth]{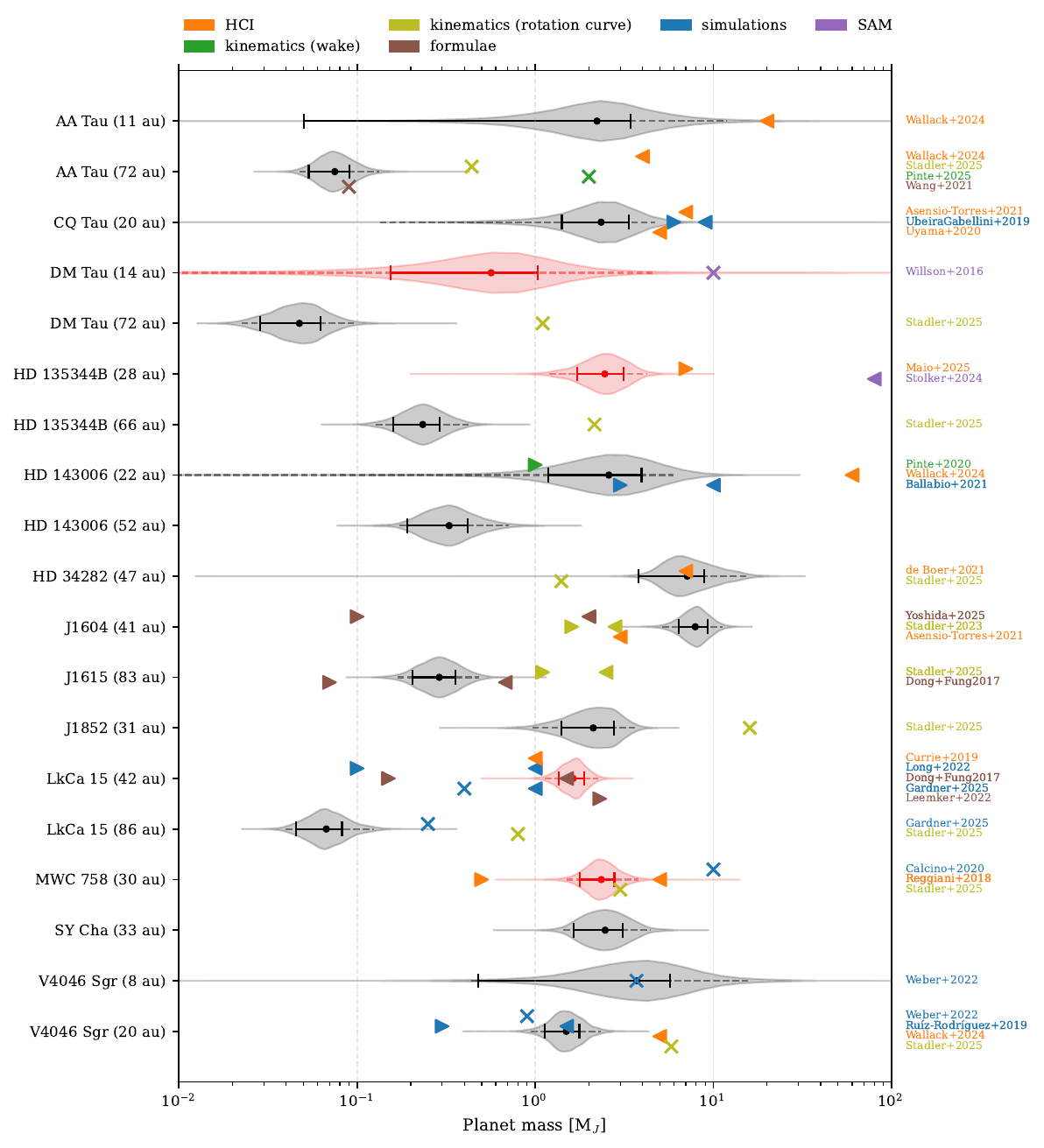}
    \caption{Overview of all inferred posterior distributions for the mass of the proposed planets (gray and red violins) and comparison with constraints provided in previous studies. The
continuous error bars mark the 16th, 50th, and 84th percentiles of the respective distribution, while the dashed lines go from the 2.5th to the 97.5th percentiles. Red violins mark uncertain estimates. Right and left facing arrows mark, respectively, lower and upper limits, while crosses indicate estimates. Different colors are used to distinguish between the methods used to provide these constraints. \emph{HCI}: high contrast imaging; \emph{SAM}: sparse aperture masking; \emph{formula}: empirical relationships with gap width and depth; \emph{simulations}: numerical modelling of dust observations; \emph{kinematics (wake)}: estimates from localized kinematic perturbations in channel maps, tracing the planet wake; \emph{kinematics (rotation curves)}: estimates from gas pressure dips observed through rotation curves. Full references can be found throughout Sect. \ref{sec:resump} and Appendix \ref{app:allsources}.}
    \label{fig:litmp}
\end{figure*}

For most of the analysed dust substructures, previous works suggested the presence of possibly embedded planets as reviewed in Appendix \ref{app:allsources}. Many of these studies put constraints on the masses of proposed planets using several approaches. Figure \ref{fig:litmp} presents a comparison of our planet mass estimates with the literature.

Non-detections in high-contrast imaging (HCI) and sparse aperture masking (SAM) observations are translated into upper limits on the planet masses. These are computed assuming different models for the planet emission and therefore depend on them and on the underlying assumptions \citep{Willson2016SparseDiscs,Reggiani2018Discovery758, Currie2019NoSignals, Uyama2020Near-infraredDisk, Asensio-Torres2021Perturbers:Disks, DeBoer2021Possible34282, Stolker2024Searching, Wallack2024ACoronagraph, Maio2025UnveilingVLT/ERIS}. We report in Fig. \ref{fig:litmp} the most constraining upper limits estimated assuming the warmer planet models, which would thus predict higher fluxes. We do not report the less constraining limits as those would, in all cases, be well above the upper end of our planet mass prior ($10^{-2}M_\star$). 
We found our estimates to be mostly consistent with these constraints, with the only two exceptions of J1604 and LkCa15 for which we estimate higher planet masses. However, results might be reconciled with colder models for the planet's emission \citep{Mordasini2012CharacterizationEvolution} or extinction from the disk material \citep[e.g.,][]{Alarcon2024ExtinctionSimulations}. In the case of LkCa15 specifically, literature studies observed significant gas emission within the cavity and argued this would be best explained by collective interaction with multiple planets \citep{Leemker2022GasCavities, Gardner2025ExoALMA.Cavity}, a situation that would be outside DBNets2.0 scope, as it was trained on examples of planet-disk interaction with only one planet.

For HD~135344~B and MWC~758, tentative detections of point-source-like emission allowed \cite{Maio2025UnveilingVLT/ERIS} and \cite{Reggiani2018Discovery758} to also put lower limits on the planet masses. Nevertheless, as stated in \cite{Maio2025UnveilingVLT/ERIS}, the presence of a CPD around the putative planet might loosen these constraints. Interestingly, the three mass estimates for the putative planets in LkCa~15 (for $R_p=42$~au), HD~135344~B and MWC~758 have a confidence score lower than the advised rejection threshold. Furthermore, all these cases, including J1604, are cavities and thus DBNets2.0 estimates can vary depending on the assumed planet position. This issue is discussed in detail in Sect. \ref{sec:rp}, illustrated with the case of J1604.

Planet masses estimated using empirical relations \citep[e.g.,][]{Dong2017WhatPlanet} with gap widths and depths and fine-tuned numerical simulations tend to agree, within the uncertainties, with DBNets2.0 results. Although J1604 and the cavity of LkCa~15 still represent notable exceptions.

Across our sample of proposed planet locations, only two of them match that of a proposed kinematic signature in disks' channel maps, discussed in \cite{Pinte2025ExoALMA.Planetsb}, for AA~Tau, and in \cite{Pinte2020NineGaps}, for HD~143006. All other kinematic perturbations proposed by \citet{Pinte2025ExoALMA.Planetsb} as planet signatures are located outside the extent of the observed continuum. 
 In the case of HD~143006, only a lower limit on the planet's mass was provided by \cite{Pinte2020NineGaps}, based on the observability of the target planet. This is consistent with our mass estimate. In the case of AA~Tau, through a comparison with synthetic data, the authors provided a planet mass estimate, which significantly overestimates our value. Interestingly, this seems common for mass estimates from kinematic signatures as discussed, for example, in \cite{Pinte2020NineGaps} and \cite{Elbakyan2022GapWinds}. With the two exceptions of HD~34282 and J1604 for which previous caveats apply, we observe the same systematic overestimation with \citet{Stadler2023ADisk, Stadler2025ExoALMA.Variations} estimates. These works provided the masses of planets potentially driving observed perturbations in the gas azimuthal velocity (rotation curves) interpreted as gas pressure dips. The planet masses were inferred using the empirical relation proposed by \cite{GyeolYun2019PropertiesDisk}, which links the planet mass to the radial width of the perturbed region. This relation was shown to exhibit weak dependence on the disk $\alpha$-viscosity.

\begin{figure*}
    \centering
    \includegraphics[width=\linewidth]{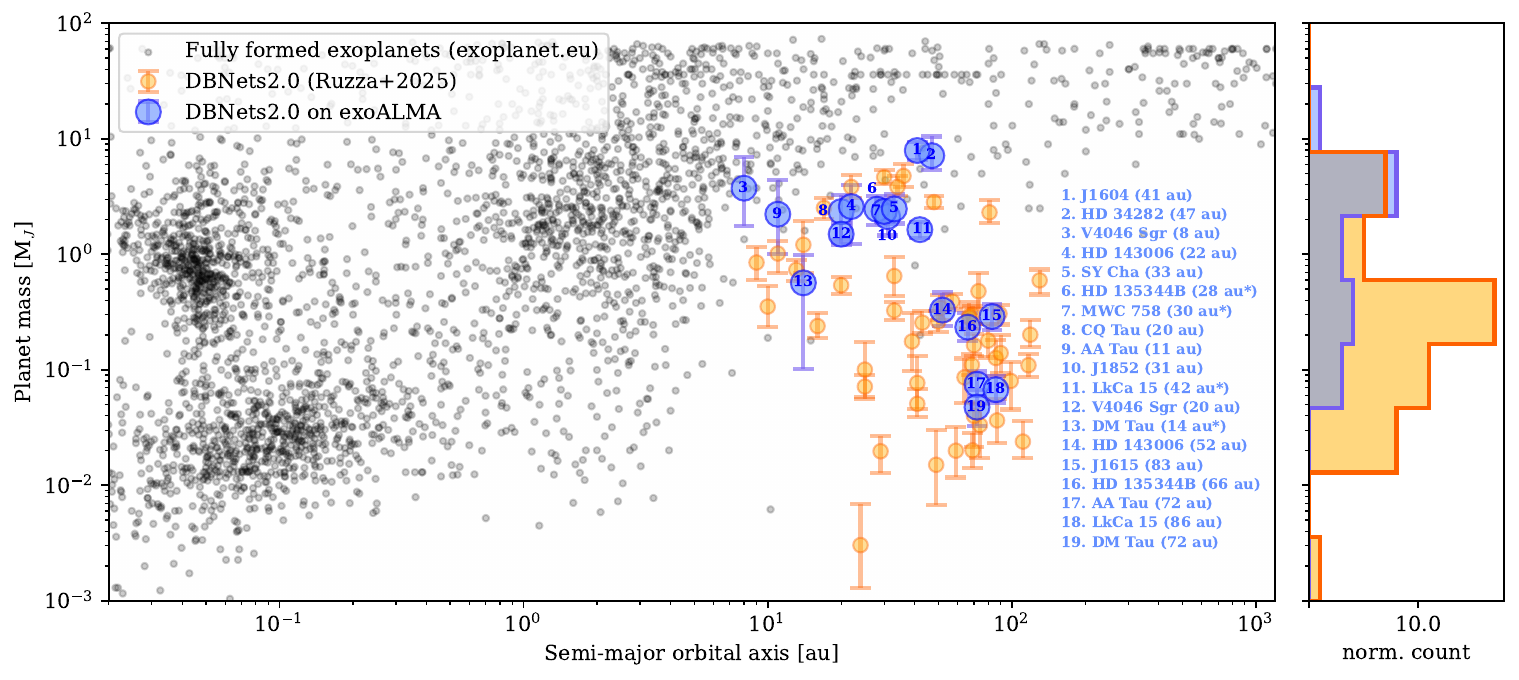}
    \caption{Inferred planet masses as a function of the putative planet's radial location. Blue points and histogram's bins mark DBNets2.0 planet mass estimates on the exoALMA sample presented in this paper. Respective disk names and assumed planet locations are listed in descending order according to the inferred planet mass. Asterisks mark uncertain or unconstrained estimates. Orange points and histogram's bins mark DBNets2.0 estimates on the larger sample of substructured disks considered in \citetalias{Ruzza2025DBNets2.0:Discs}.
    Black points mark all confirmed exoplanets (data from \href{https://exoplanet.eu}{exoplanet.eu}).}
    \label{fig:ma}
\end{figure*}

Figure \ref{fig:ma} shows the masses of the proposed planets inferred using DBNets2.0 as a function of their putative location. It is remarkable how proposed planets in dust substructures populate a region of this parameter space that was otherwise inaccessible with standard exoplanet detection techniques, and suggests the possibility of different planetary system architectures. On the other hand, this planet's population would also be significantly younger, raising the question of how it would evolve with time. Furthermore, looking at individual sources in this plot is particularly relevant for selecting suitable targets for direct imaging of the planet's thermal emission, which is easier to observe in the case of massive planets far from the host star. According to these criteria, our analysis suggests HD34282 and J1604 as the most promising candidates. This remains true even including the substructured disks analyzed in \citetalias{Ruzza2025DBNets2.0:Discs}. We should note, though, that this result refers to putative planets in the specific locations indicated in Table \ref{tab:allres} and informed by previous studies. As discussed in Sect.~\ref{sec:rp}, these results could change if the planet orbits were different.
In the right panel of the same figure we show that proposed planets in the substructures of the exoALMA disk sample are typically more massive than those emerging from the larger sample of dust substructures analyzed in \citetalias{Ruzza2025DBNets2.0:Discs}.

\subsection{Turbulent viscosity}
\label{sec:dmtau}

As mentioned in Sect. \ref{sec:results}, the distribution of $\alpha$-viscosity for the entire sample is skewed towards the lower end of the DBNets2.0 prior.
DM Tau is one of the few cases for which there are robust measures of line broadening interpreted as a high level of turbulence \citep{Guilloteau2012ChemistryDisks, Flaherty2020MeasuringSgr} corresponding to $\alpha=0.08 \pm 0.02$. Similarly, \citet{C.Hardiman2025ExoALMADisk} performed a full radiative transfer modelling of \ce{^12CO}~J=3--2 and \ce{CS}~J=7--6 finding that data are best fit by non-thermal line broadening corresponding to $\alpha=0.16^{+0.01}_{-0.02}$. It must be noted that our training of DBNets2.0 with synthetic data with $\alpha$ values up to $10^{-2}$ effectively sets a prior on the returned estimates, preventing the inference of higher values. Nevertheless, the DBNets2.0 fit of dust substructures in this disk yields a picture consistent with that in the presented literature. 

For the gap at 14~au $\alpha$ is not constrained by our analysis, probably due to the low spatial resolution, as the obtained posterior extends over the entire support of the prior. The analysis of the gap at 72~au instead returns an estimate $\alpha=5.5^{+3.1}_{-2.0} \times 10^{-3}$, skewed towards higher values than the other disks. Although this is not directly compatible with the mentioned literature estimates, several caveats apply. First, as already mentioned, DBNets2.0 priors are too constraining in this case; second, line broadening measures refer to the localized regions where the molecular emission is detected, which are both at larger radii and well above the midplane, whereas our estimate is sensitive to the turbulence near the midplane at the substructure radial location. Modeling indeed indicates that, depending on the mechanism, velocity perturbations could increase with height \citep[see, e.g.][for MRI-driven turbulence]{Simon2015SIGNATURESOBSERVATIONS}.

\subsection{Aspect ratios and Stokes numbers}

\begin{figure}
    \centering
    \includegraphics[width=0.9\linewidth]{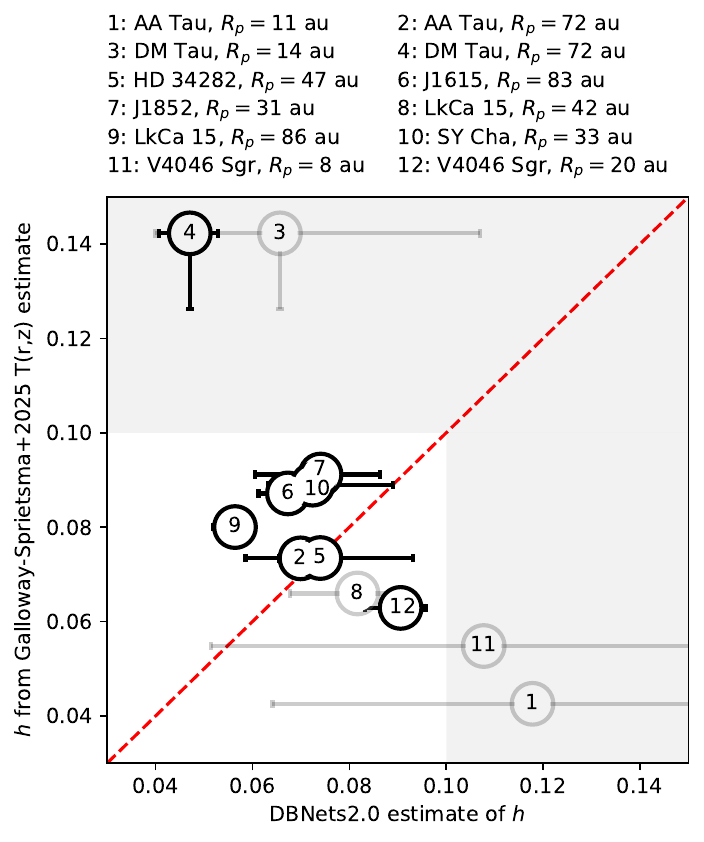}
    \caption{Comparison between the aspect-ratio estimates obtained with DBNets2.0 and the temperature-profile measurements reported by \cite{Galloway-Sprietsma2025ExoALMA.Structures}.
For this comparison, the midplane temperatures are converted into disk aspect ratios under the assumptions of vertical hydrostatic equilibrium and a vertically isothermal structure, consistent with the physical model used to train DBNets2.0. Faded points indicate sources with unreliable or unconstrained estimates. The red dashed line denotes perfect agreement between the two methods. The gray background marks parameter values that fall outside of DBNets2.0's scope.}
    \label{fig:comparisonH}
\end{figure}

Via line profile analysis of exoALMA observations, \citet{Galloway-Sprietsma2025ExoALMA.Structures} located the emission surfaces of observed emission lines, deriving the thermal structures of some exoALMA disks. These temperatures, extrapolated to the disks' midplanes, can be compared to our estimates of the disks' aspect ratios by converting them under the assumption of vertical hydrostatic equilibrium. Figure \ref{fig:comparisonH} presents this comparison done by using \cite{Galloway-Sprietsma2025ExoALMA.Structures} estimates of the midplane temperature profile to compute the expected aspect ratio at the putative planet locations where DBNets2.0 estimates are given. We observe no correlation between the two estimates, although \citet{Galloway-Sprietsma2025ExoALMA.Structures} typically found higher temperatures. The same was observed and discussed in \citet{Galloway-Sprietsma2025ExoALMA.Structures} comparing their temperature estimates with those derived from disks' scale heights and simulations, suggesting the need for an improved temperature profile prescription. Although in some cases \citet{Galloway-Sprietsma2025ExoALMA.Structures} temperatures imply disks aspect ratios higher than those in DBNets' prior, we note (see Fig. \ref{fig:allsamp_hist}) that our estimates do not saturate towards the prior's upper limit.

Regarding Stokes number estimates, we highlight the case of LkCa~15. For this disk, \citet{Sierra2025HighLkCa15} performed a multiwavelength analysis of the observed rings, deriving, through a comparison with dust trapping models, a plausible range of dust Stokes numbers for each of the three rings at 42, 69, and 101 au. However, these also depend on the chosen model for the opacity. With DBNets2.0, we were only able to constrain St close to the gap at 86 au. In this case, we estimate $\log \text{St} = -1.45^{+0.12}_{-0.14}$. This is in line with \citet{Sierra2025HighLkCa15} estimates for the two outer rings \mbox{$-1.17 < \log \text{St} < 0.36$} that assume DSHARP opacities \citep{Birnstiel2018TheModel}. Using \citet{Ricci2010DustWavelengths} opacities instead yielded the wider range $-2.46 < \log \text{St} < 0.81$, which covers almost entirely the DBNets2.0 prior (see Appendix \ref{app:scope}).

\section{Discussion}
\label{sec:discussion}

\subsection{Dependence of DBNets2.0 estimates on the assumed planet location}
\label{sec:rp}
\begin{figure}
    \centering
    \includegraphics[width=\linewidth]{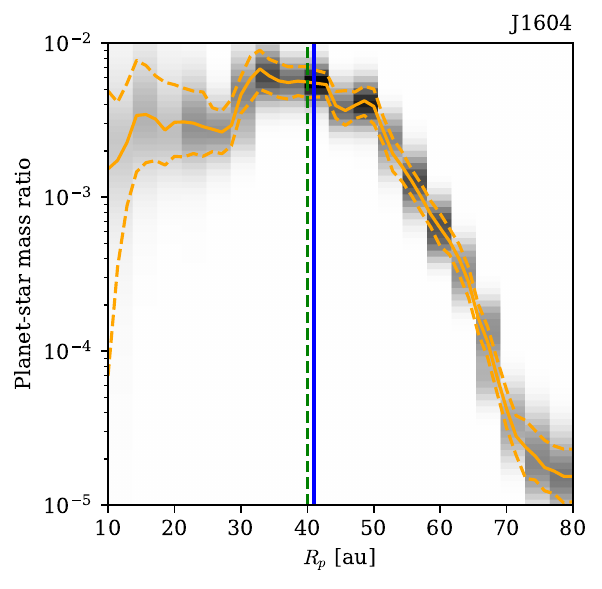}
    \caption{Degeneracy of DBNets2.0 planet mass estimate with the assumed planet radial location in the cavity of the J1604 disk. The overlayed orange lines mark the 16th, 50th and 84th percentiles of the inferred distribution. The vertical blue line marks the location of the putative planet assumed in this study. The vertical green dashed line indicates $R_\text{edge}/2$, where $R_\text{edge}$ is the radial location of the peak of the ring outside the cavity. }
    \label{fig:rmpdeg}
\end{figure}

We explained in Sect. \ref{sec:dbnets20} that the radial location of the planets proposed in disk substructures must be provided as one of DBNets2.0 inputs. This is needed to rescale the observation and match the physical scale of the training dataset. 

The inference results can easily change with different assumptions on the planet location as a natural consequence of the change, after rescaling, of measures of substructures' morphological features, such as gaps and rings widths, as well as their distance to the planet.
This ambiguity is not particularly relevant for the analysis of gaps, as the alleged planet location can be easily constrained, for example, assuming the intensity minimum in the azimuthally averaged radial profile.
For cavities, instead, the situation is more delicate as there is often no indication of plausible planet locations and the range of possible positions is typically wide. 
Figure \ref{fig:rmpdeg} illustrates this degeneracy for the inferred planet mass estimates in the specific case of the J1604 cavity terminating with a bright ring at 80~au. As expected, the inferred planet mass decreases as the assumed planet location approaches the ring.

The same is expected for all other planets assumed within disk cavities. For this reason, in those cases, we only included in our sample planet locations previously suggested by independent methods and observations.
We present plots similar to Fig. \ref{fig:rmpdeg} for the other inferred disk properties in Appendix \ref{app:rpdeg} along with the same analysis performed for all the cavities in our sample.
We define the cavity radius identifying $R_\text{edge}$ as the peak of the ring outside the cavity. In general, the estimated planet mass appears independent of the assumed radial location as long as it lies within half the cavity radius (green dashed line in Fig. \ref{fig:rpdeg} and \ref{fig:rpdeg2}), while it decreases progressively as the planet approaches the outside ring. As shown in the same figure, estimates for the other disk properties are also affected by the choice of the planet position, although their dependence on it is less intuitive. 

\subsection{Comparison with DBNets2.0 estimates in \citet{Ruzza2025DBNets2.0:Discs}}
\label{sec:compR25}
\begin{figure}
    \centering
    \includegraphics[width=\linewidth]{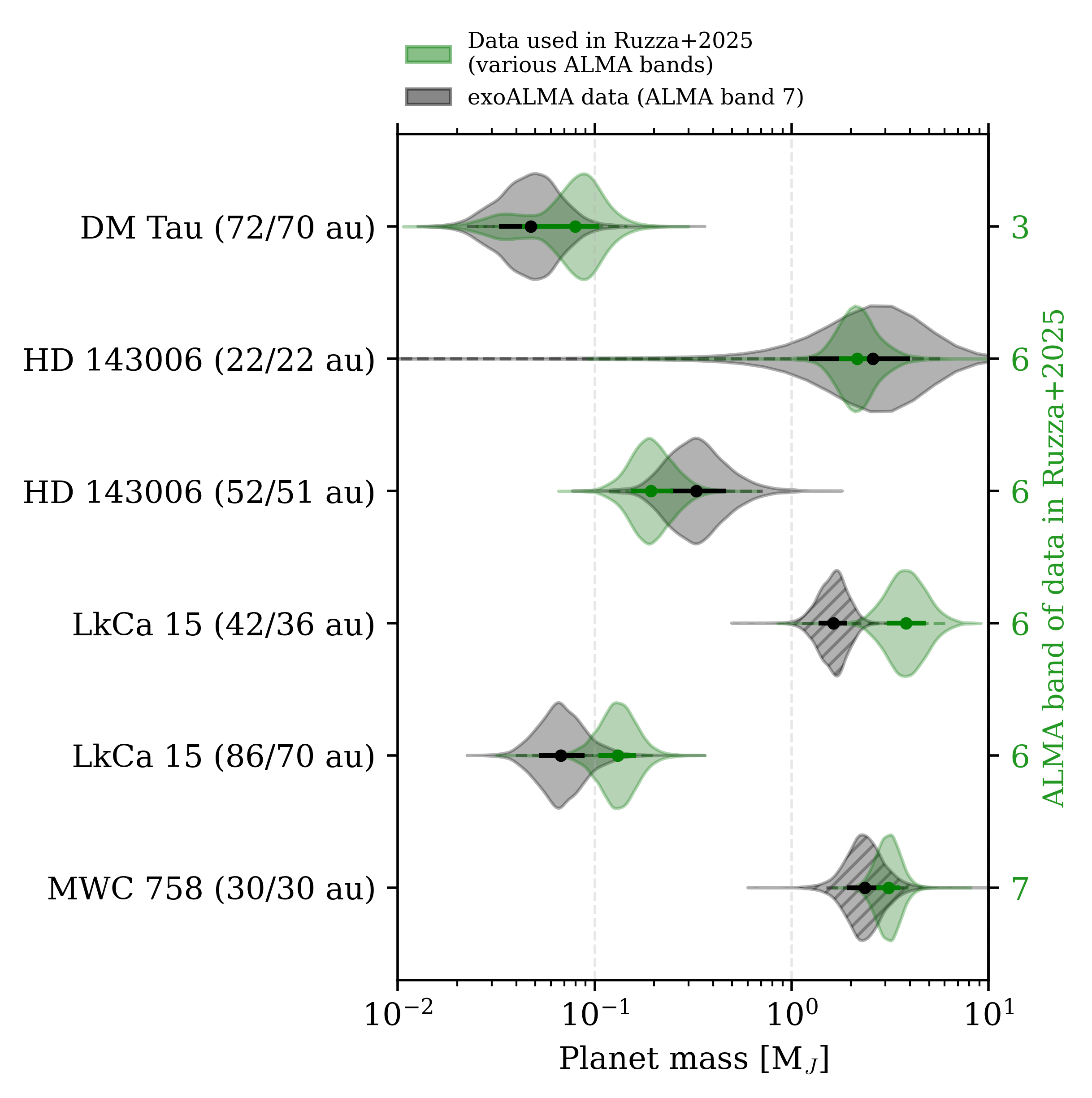}
    \caption{Comparison, for the overlapping objects, of DBNets2.0 estimates derived in \citetalias{Ruzza2025DBNets2.0:Discs} on archival data (green violins) with those obtained in this work on the exoALMA observations (black violins). On the right, in green, the Band of the continuum observation used in \citetalias{Ruzza2025DBNets2.0:Discs}. Near each disk name we report the assumed planet locations, respectively in \citetalias{Ruzza2025DBNets2.0:Discs} and in this work. Violin plots of estimates corresponding to confidence scores below the acceptance threshold in either work are hatched.}
    \label{fig:comp_ruzza25}
\end{figure}

In Fig. \ref{fig:comp_ruzza25} we compare, for the overlapping sample, DBNets2.0 inferred planet masses on the exoALMA Band~7 ($\lambda=0.8–1.1$~mm) data with those presented in \citetalias{Ruzza2025DBNets2.0:Discs} for the same objects but obtained from the analysis of different observations, in most cases taken at different wavelengths (band 6: $\lambda=1.1–1.4$~mm; band 3: $\lambda=2.6–3.6$~mm). Interestingly, estimates are compatible within uncertainties, proving little dependence of these results on the Band of the observations. We separately highlight the case of MWC~758 for which we are comparing the analysis of observations at the same wavelength but with different resolutions. Specifically, the observation used in \citetalias{Ruzza2025DBNets2.0:Discs} is better resolved by a factor of 2, which translates to a compatible but more precise estimate than that obtained on the exoALMA data. The same is also observed across different wavelengths for the case of HD~143006, for which the observation used in \citetalias{Ruzza2025DBNets2.0:Discs} is resolved approximately 2.7 times better.

This general agreement is also observed for the other inferred properties, as shown in Fig.~\ref{fig:compR25} of Appendix~\ref{app:compr25}. The largest difference emerges for the inferred $\alpha$-viscosity in DM~Tau, for which the DBNets2.0 analysis of a longer wavelength (Band 3) observation results in a lower value by more than one order of magnitude. Roughly speaking, a continuum observation mainly traces dust grains of size $a=\lambda/2\pi$, where $\lambda$ is the observed wavelength. Therefore, since $St \propto a$, a change $\Delta \lambda /\lambda$ should correspond to the same relative change $\Delta \text{St}$ in the Stokes number of the observed dust grains. This relative difference is $\sim 30$\% for bands 6/7 and $\sim 60\%$ for bands 7/3. \citetalias{Ruzza2025DBNets2.0:Discs} demonstrated that DBNets2.0 tipically achieves a 10\% precision on the Stokes number estimates and thus should be sensitive to a change in observational wavelength. This, however, is not observed in practice, probably because the effect of St on the rings and gaps morphology is degenerate with $\alpha$. We already noted this in \citetalias{Ruzza2025DBNets2.0:Discs}, where we interpreted it as indicating that the inference is primarily driven by morphological features associated with dust trapping and diffusion (e.g. rings' prominence).  Hence, DBNets2.0 may be interpreting the differences in dust substructures as due instead to a different $\alpha$-viscosity, which indeed varies the most when comparing Band 3 and 7 observations. Furthermore, the fact that DBNets2.0 was trained on synthetic observations generated at Band 6 wavelengths might play a role.
From this analysis, we conclude that we should not fully trust DBNets2.0 estimates on $\alpha$ and St when those are inferred using observations at wavelengths significantly different than those of Band 6. Figure \ref{fig:comp_ruzza25} and \ref{fig:compR25} instead show that using Band 7 observations yields compatible results within the given uncertainties.

\subsection{Accretion timescales}

\begin{figure*}
\centering
\includegraphics[width=\linewidth]{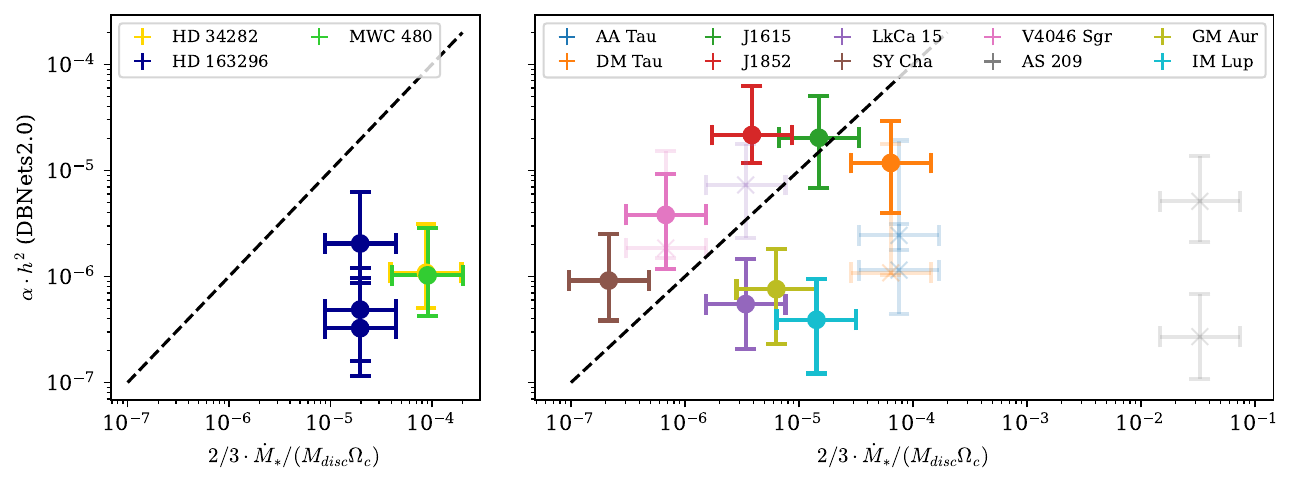}
\caption{Comparison of DBNets2.0 local estimates of viscous-accretion timescales with the global accretion rates derived from measurements of star accretions \citep{Donati2011TheAccretion,Fairlamb2015ARates, Manara2023DemographicsFormation} and disk masses \citep{Martire2024RotationDisks,Longarini2025ExoALMA.Modelling}. 
The left panel presents the comparison for all Herbig stars in the sample, while the right panel refers to T Tauri stars.
Faded points mark unreliable or unconstrained estimates of $\alpha h^2$ or $M_\text{disk}$. The dashed black lines mark the equality between the quantities on the two axes.}
    \label{fig:ah2disc}
\end{figure*}

Assuming a self-similar density profile, \cite{Longarini2025ExoALMA.Modelling} modelled the rotation curves of a subsample of exoALMA targets, fitting the stellar mass, disk mass, and scale radius with the method introduced in \citet{Lodato2022DynamicalDiscs}. The same analysis was also performed on the MAPS \citep{Oberg2021MoleculesHighlights} disks sample in \cite{Martire2024RotationDisks}. Combining these results with literature measures of the stars' accretion rates, \cite{Longarini2025ExoALMA.Modelling} constrained the disk accretion timescales, assuming the self-similar solution to the disk evolution, as \citep{Hartmann2016AccretionStars}
\begin{equation}
\label{eq:acctime_ss}
    (t_{\text{acc}} / t_\text{dyn}) ^ {-1} \sim \alpha h^2 = \frac{2}{3} \frac{\dot M}{M_\text{disk} \Omega_c}
\end{equation}
where $t_\text{dyn} = \Omega^{-1}$ and $\Omega_c$ is the Keplerian velocity at the disk scale radius.
This relation captures the global accretion flow and, as cautioned by \cite{Longarini2025ExoALMA.Modelling}, the derived $\alpha$ (or $\alpha h^2$) must be interpreted as an estimate of the disk accretion regardless of the mechanism that drives it.
Conversely, with DBNets2.0, we can provide local estimates of $\alpha h^2$ obtained by modelling substructures due to planet-disk interaction, whose formation and morphology are sensitive to the value of viscosity (or turbulence). 

In the scatter plots of Fig. \ref{fig:ah2disc} we compare the estimates of the two methods for the exoALMA disks for which dynamical masses have been measured \citep{ Longarini2025ExoALMA.Modelling}. We combine in the same plot the same estimates for MAPS disks, obtained using \cite{Martire2024RotationDisks} disks dynamical masses and \citetalias{Ruzza2025DBNets2.0:Discs} DBNets2.0 estimates for $\alpha$ and $h$. 
The left panel presents these results for Herbig stars in the considered sample, while T Tauri stars are shown in the right panel. 
Interestingly, we observe that for all the Herbig stars in our sample, our measure of the disk timescale for viscous transport cannot explain the observed accretion rate, suggesting that additional accretion mechanisms should be invoked to reproduce the observed timescales.
The results obtained for T Tauri stars present instead a larger scatter, although in most cases the two estimates are roughly consistent, implying that the magnitude of viscous transport inferred from planet-disk interaction signatures could alone explain the measured accretion rates. Note that we marked with fading colors all points for which one of the two estimates is considered uncertain or unconstrained.
DBNets2.0 overestimation of $\alpha h^2$ for some disks may be due to their substructured nature, which would alleviate the validity of the self-similar hypothesis and, therefore, of equation \ref{eq:acctime_ss}. Indeed, some studies suggested that the presence of planets could reduce mass accretion into the host star (e.g. \citealt{MullerTobiasW.A.2013ModellingDisks, ManaraC.F.2019ConstrainingRates, Bergez-CasalouC.2020InfluenceDiscs})

\subsection{Kanagawa coefficients and stalling radius}
\begin{figure}
    \centering
    \includegraphics[width=\linewidth]{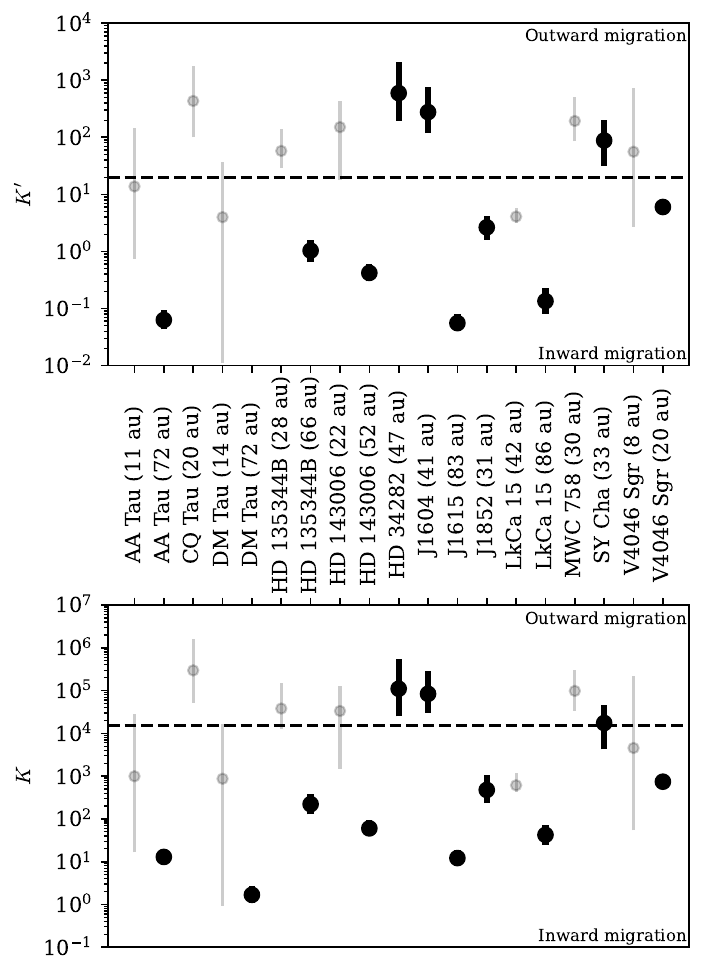}
    \caption{The $K'$ (upper panel) and $K$ (lower panel) coefficients, defined in Eq. \ref{eq:kp} and \ref{eq:k}, derived for all analyzed disks from DBNets2.0 estimates.  The horizontal dashed lines mark the threshold values of $K'=20$ and  $K=1.5\cdot 10^4$ for the expected direction of planet migration.
    Faded points indicate sources with unreliable or
unconstrained estimates.}
    \label{fig:kcoeff}
\end{figure}

The effects and physics of planet-disk interaction and their dependence on the system properties are often summarized into two parameters $K$ and $K'$ defined as:
\begin{align}
\label{eq:k}
    K' &= \left(\frac{M_p}{M_\star}\right)^2 \alpha^{-1}h^{-3}, \\
    \label{eq:kp}
    K &= \left(\frac{M_p}{M_\star}\right)^2 \alpha^{-1} h^{-5}.
\end{align}
For example, \cite{Kanagawa2015FormationRotation} have found that the depths of the gas gaps carved by planets scale approximately as $1/(1+0.04K)$, while \cite{Kanagawa2016MassWidth} showed that the gap width scales as $\Delta_\text{gap}~\propto~K'^{1/4}$.
In Fig. \ref{fig:kcoeff} we show the $K$ and $K'$ computed for each proposed planet using DBNets2.0's estimates of the disk and planet properties.

An interesting application of these parameters regards the phenomenon of planet migration for which numerical modelling \citep{ Dempsey2021OutwardSuper-Jupiters, Scardoni2022InwardRadius} has shown $K$ and $K'$ values to control its direction (inward or outward). Specifically, as their values increase, at the thresholds $K=1.5\cdot 10^{4}$ and $K'=20$, the planet would switch from inward to outward migration.
The values shown in Fig. \ref{fig:kcoeff} indicate that, excluding points with unconstrained or uncertain estimates and independently of assuming migration to be determined by $K$ or $K'$, most of our planets would fall into the inward migrating case, with only three objects migrating outwards.

Typically, as the disk aspect ratio is expected to increase further from the star, both $K$ and $K'$ would decrease at increasing radii, thus an initially outward migrating planet would at some point switch to inward migration. Following this reasoning \cite{Scardoni2022InwardRadius} proposed the existence of a stalling radius which would correspond to the locations where $K$ and $K'$ equal the threshold values controlling the migration sign. 
The high scatter in the $K$ and $K'$ inferred values indicates that we do not observe an accumulation of planets in proximity of the expected stalling radii.

It is important to note that the model used to analyze dust substructures, implicitly defined by the simulations used to train DBNets2.0 algorithms, does not include planet migration, as planets are held on a fixed orbit. Therefore, the effects of this phenomenon on the dust morphology \citep{Nazari2019RevealingObservations, Meru2019IsRings} are not considered.

\subsection{Configurations of proposed multiple systems}
\begin{figure}
    \centering
    \includegraphics[width=\linewidth]{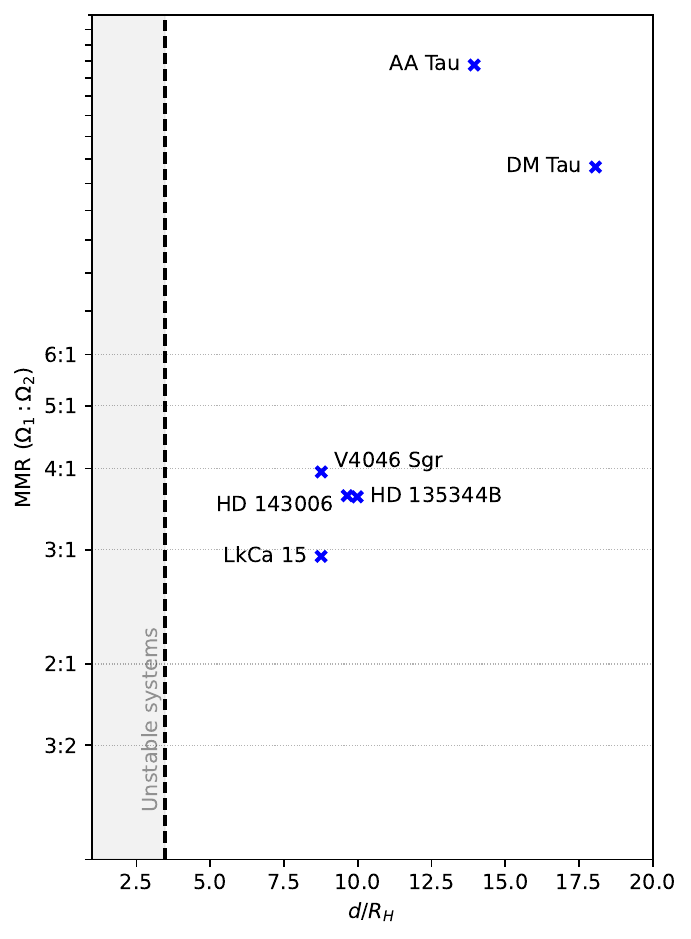}
    \caption{Disks where we assumed two fiducial planet locations. The x-axis shows the mutual distance of the two planets in the proposed three-body systems with respect to their mutual Hill radius. The vertical dashed line marks the stability threshold $d/R_H = 2\sqrt 3$ found by \cite{Gladman1993DynamicsPlanets}. According to their criterion, orbits of planets whose mutual distance is lower are unstable. The y-axis indicates the ratio between the keplerian angular velocities at the two planet locations. Ticks on the y-axis indicate mean motion resonances (MMR).}
    \label{fig:stability}
\end{figure}

For six disks in the sample, we propose the presence of two planets. We thus question first whether the proposed planet locations correspond to specific resonances and, second, the stability of these three-bodies systems. We check the latter according to the Hill criterion proposed by \cite{Gladman1993DynamicsPlanets}, which poses a threshold on the mutual distance of the two planets $d$ in terms of their mutual Hill radius $R_H = [(m_1+m_2)/3M_\star]^{1/3}[(a_1+a_2)/2]$. This is required to be greater than $2\sqrt{3}$ for the system to be stable. Results are presented in Fig. \ref{fig:stability} where, for each one of the six proposed multiple systems, the x-axis indicates the mutual distance of the two planets over their mutual Hill radius $d/R_H$ while the y-axis is the ratio between their Keplerian velocity. According to this analysis, all these systems would 
be dynamically stable and are not close enough to lie in the lower-order mean-motion resonances (such as 3:2, 2:1).

\subsection{Correlations with other disk properties}
We explored any possible relation between DBNets2.0 best estimates for $\alpha, h, \text{St}, \text{and } M_p$ and disk properties measured on the exoALMA data. Specifically, we considered the disk gas-to-dust mass ratios, the non-axisymmetry indexes quantifying the axisymmetry of continuum observations , the disk dynamical masses, the disk dust masses, and the host star masses \citep{Curone2025ExoALMA.Emission, Longarini2025ExoALMA.Modelling, Izquierdo2025ExoALMA.Disks}. We do not observe any statistically significant correlation, with the sole exception of the pair stellar mass-aspect ratio, which is detailed in the right panel of Fig. \ref{fig:correlations}. This likely arises because more massive stars have higher luminosity, and thus their stronger irradiation is reflected in the disk temperature. Details and results of this analysis are presented in appendix \ref{app:correlations}. 

\section{Conclusion}
\label{sec:conclusion}

In this paper, we examine the dust substructures revealed by exoALMA continuum observations, interpreting them as signatures of planet-disk interaction to characterize the properties of possible embedded planets and their host disks. For this purpose, we employed the SBI pipeline DBNets2.0 \citepalias{Ruzza2025DBNets2.0:Discs} which exploits deep learning methods to quickly and accurately perform a statistical inference of the disk $\alpha$-viscosity, scale height, dust Stokes number, and planet mass. The inference of full posterior distributions allowed for proper uncertainty quantification, and the application of two criteria made it possible to flag and separate uncertain estimates, ensuring the robustness of the overall results.

Carefully inspecting each observation and reviewing previous studies we propose a population of 19 planets possibly embedded in 13 of the 15 disks targeted by the exoALMA program. After applying DBNets2.0 to each one of these putative planets we discuss the results, their comparison with previous works, and their implications.
The main outcomes of this study are the following:
\begin{enumerate}

    \item We identify dust substructures in the exoALMA disk sample that might most promisingly host young planets. We provide mass estimates for these putative planets and discuss them in the context of the current literature. Most of our estimates are consistent with upper limits set by high-contrast imaging and estimates from simulations or empirical formulae. Putative planets in the cavities of J1604 and LkCa15 are expected to be more massive than what constrained by previous estimates. However, the estimate for LkCa15 is associated with a low confidence score and, in both cases, these estimates depend on the assumed planet location. Specifically, a lower mass would be consistent with the planet being further from the star.
    
    \item Due to their separation from the host star and inferred high mass, we find the proposed planets in HD34282 and J1604 to be the most suitable for attempting direct imaging observations.
    
    \item Most of the proposed planets are super-Jovian, likely due to sample selection bias.
    
    \item Cavities require larger planet masses than gaps.
    
    \item Interpreting the observed substructures as due to planet-disk interaction, we provide estimates of 3 disk properties: $\alpha$-viscosity, disk scale height and dust Stokes number. For DMTau our analysis interestingly suggests a high value of $\alpha$-viscosity, in agreement with constraints from line broadening measures \citep{Guilloteau2012ChemistryDisks, Flaherty2020MeasuringSgr}. 
    
    \item Derived estimates of the disks' viscous timescales ($\alpha h^2$) are too long to explain the measured accretion rates of Herbig stars, while they present a trend consistent with most measures of accretion rates and disk masses in TTauri stars.
    
    \item Derived parameters controlling the migration direction indicate the regime of inward migration to be more frequent, although three putative planets would undergo outward migration.
    
    \item We find a statistically significant correlation between the stellar masses and inferred disks' aspect ratio (Pearson correlation coefficient of 0.60). We do not observe any other correlation of the inferred quantities with other disk properties. 
   
\end{enumerate}

In addition to these results, the analysis demonstrates the effectiveness of SBI tools, and in particular DBNets2.0, for performing quick and systematic studies of large observational surveys. Homogeneous samples of disk and planet properties, affected by the same model and observational assumptions and systematics, can help address fundamental questions of planet formation and evolution, as demonstrated in this work. While not all dust substructures may originate from planet-disk interactions, if a substantial fraction do, then systematic analyses of large samples can still provide meaningful insights into the population-level properties of young planets, especially when combined with complementary observations and methods. Single results can additionally explain the outcomes of direct imaging surveys and lead future observations towards the most promising candidates.

\begin{acknowledgments}
This paper makes use of the following ALMA data: ADS/JAO.ALMA\#2021.1.01123.L. ALMA is a partnership of ESO (representing its member states), NSF (USA) and NINS (Japan), together with NRC (Canada), MOST and ASIAA (Taiwan), and KASI (Republic of Korea), in cooperation with the Republic of Chile. The Joint ALMA Observatory is operated by ESO, AUI/NRAO and NAOJ. The National Radio Astronomy Observatory is a facility of the National Science Foundation operated under cooperative agreement by Associated Universities, Inc.

GR acknowledges support from the Fondazione Cariplo,
grant n° 2022-1217 and from the European Research Council (ERC) under grant agreement no. 101039651 (DiscEvol). GL acknowledges support from PRIN-MUR 20228JPA3A, from the European Union Next Generation EU, CUP:G53D23000870006. GL and AR acknowledge support from the European Union’s Horizon Europe Research \&
Innovation Programme under the Marie Sklodowska-Curie grant agreement No.
823823 (DUSTBUSTERS).
S.F. is funded by the
European Union (ERC, UNVEIL, 101076613) and acknowledges financial contribution from PRIN-MUR 2022YP5ACE.
J.B. acknowledges support from
NASA XRP grant No. 80NSSC23K1312.
M.B., J.S., and D.F. have received funding
from the European Research Council (ERC) under the
European Union’s Horizon 2020 research and innovation
program (PROTOPLANETS, grant agreement No.
101002188).
PC acknowledges support by the ANID BASAL project FB210003.
C.H. acknowledges support from NSF AAG grant No. A. T.H.
is supported by an Australian Government Research Training
Program (RTP) Scholarship. Support for A.F.I. was provided by NASA
through NASA Hubble Fellowship grant No. HST-HF2-
51532.001-A awarded by the Space Telescope Science
Institute, which is operated by the Association of Universities
for Research in Astronomy, Inc., for NASA, under contract
NAS5-26555.
C.L. has received funding from the UK Science and
Technology research Council (STFC) via the consolidated
grant ST/W000997/1. C.P.
acknowledges Australian Research Council funding via
FT170100040, DP18010423, DP220103767, and
DP240103290.   A.J.W. has been supported by the Royal Society through a University Research Fellowship, grant number URF\textbackslash R1\textbackslash 241791. 		
T.C.Y. acknowledges support by grant-in-aid for JSPS Fellows
JP23KJ1008. Support for B.Z. was provided by the Brinson
Foundation.
FMe has received funding from the European Research Council (ERC) under the European Union's Horizon Europe research and innovation 
program (grant agreement No. 101053020, project Dust2Planets). Views and opinions expressed by ERC-funded
scientists are, however, those of the author(s) only and do not
necessarily reflect those of the European Union or the
European Research Council. Neither the European Union nor
the granting authority can be held responsible for them.
\end{acknowledgments}

\appendix

\section{Details of DBNets2.0 and underlying assumptions}

\label{app:dbnetsmodel}
As specified in Sect. \ref{sec:dbnets20}, DBNets2.0, as any other SBI model, is ultimately fitting the input data with a model that is defined by its training dataset. This consists of protoplanetary disks' synthetic observations of the dust continuum emission with planet-induced substructures. To generate these synthetic observations, we first ran hydrodynamical simulations and then postprocessed their results. In this appendix, we provide details on this procedure and on additional aspects of DBNets2.0. We highlight the main underlying assumptions and briefly discuss their implications for interpreting the results. Further details can be found in \citetalias{Ruzza2025DBNets2.0:Discs}.

\subsection{Details on the hydrodinamical simulations}
Hydrodynamical simulations are run with the mesh code FARGO3D (\citealt{Benitez-Llambay2016FARGO3D:CODE}, see \citealt{Bae2025ExoALMA.Codes} for a systematic comparison between hydrodynamical codes).
One, non-accreting, planet is embedded in the disk and maintained on a fixed orbit. We adopt a locally isothermal equation of state with a set temperature profile. 
The disk evolves viscously with viscosity parametrized through \cite{Shakura1973BlackAppearances} prescription with constant $\alpha$ across the entire simulation domain; this effective viscosity is used to model the effects of turbulence. 
Dust is modelled as a pressureless fluid subject to diffusion, with Schmidt number $Sc=1$, and gas-drag with fixed Stokes number across the entire simulation domain. Back reaction of the dust on the gas is neglected.

\subsection{Radiative transfer and additional postprocessing of synthetic observations}

Using the simulated disks dust density ($\Sigma_d$) and temperature ($T_d$), synthetic observations are generated by computing the surface brightness as $T_s = T_d(1 -e^{\kappa \Sigma_d})$, which is the radiative transfer solution for a vertically homogenous (in temperature and dust properties), face-on, disk. The opacity ($\kappa$) is computed using \cite{Birnstiel2018TheModel} model.
We note that other ML-based methods \citep[e.g.,][]{ShafaatMahmud2025InferringAutoencoder} exist that rely on radiative transfer calculations. In our setup (based on 2D simulations with a prescribed temperature structure and a single-sized dust grain population) the primary advantage of radiative transfer methods is their ability to properly account for disk inclination, which causes photons to traverse longer paths spanning a range of disk radii. We expect this effect to introduce significant differences only for highly inclined disks, which is not the case for the sample considered here.
During training, Gaussian noise with random variance is added, and the images are convolved with Gaussian beams of varying sizes to ensure that the results are robust to these observational effects.

\subsection{Training set parameter space (effectively setting the inference priors)}
\label{app:scope}
The training set covers the parameter space region where $10^{-4}\leq\alpha\leq10^{-2}, 0.03\leq h \leq0.1, 10^{-3} \leq \text{St} \leq 10^{-1}, 10^{-5} \leq M_p/M_\star \leq 10^{-2}$. This effectively sets uniform priors for each of these inferred properties in the respective range of values.
Two additional parameters, $\sigma$ and $\beta$ are varyied, within our training dataset. These are the slopes of the power laws that set the initial radial profile of, respectively, the disk surface density and aspect ratio. The value of $\beta$ is uniformly sampled in the range $0 \leq \beta \leq 0.35$ while $\sigma$ is set to match the steady state solution as $\sigma=2\beta + 1/2$, therefore spanning the range $0.5 \leq \sigma \leq 1.2$. DBNets2.0 inferred posteriors are marginalized with these priors over these parameters.

\subsection{Definition of the confidence score}
 
The confidence score, introduced in section \ref{sec:dbnets20}, given an observation $x$ and DBNets2.0 estimate $p(\theta|x)$, is computed as follows: 1) we sample ten $\theta_i$ from $p(\theta|x)$, 2) we linearly interpolate the synthetic images in the training dataset at all $\theta_i$, 3) we remap the interpolated images $b_i$ and the input image $x$ to polar coordinates, 4) we compute the FFT of the remapped $b_i$ and $x$, 5) we compute CS as 
\begin{equation}
    \text{CS} = 1 - \frac {1} {10} \sum_i \frac{(|F(x)| - |F(b_i)|)^2 }{(|F(x)|^2 + |F(b_i)|^2)}.
\end{equation}

\subsection{Deprojection and preprocessing of input observations}
Observations need to be deprojected before being input into DBNets2.0. We do that by linearly interpolating the observation and using the geometrical properties fitted by \cite{Curone2025ExoALMA.Emission}. In \citet{Ruzza2024DBNets:Discs} we proved, on synthetic data, that results remain unaffected by deprojection effects up to a disk inclination of 60°. The result should also hold for \citetalias{Ruzza2025DBNets2.0:Discs}, as both build on the same CNN to process the data. All disks in the exoALMA sample satisfy this criterion. We note that the CS is computed using the input image after deprojection. Therefore, possible artifacts introduced by the deprojection could, in principle, lead to a reduction in CS, although this has not been tested.

Input observations are also standardized by subtracting the mean value of their pixels and dividing by their standard deviation \citepalias[see][]{Ruzza2025DBNets2.0:Discs}.

\subsection{Treatment of multiple substructures}
For the analysis of disks with multiple substructures, we separately consider each gap as a viable planet location and independently fit our single-planet model. Each DBNets2.0 output considers, in principle, the entire observed morphology within $0.3R_p$ and $4R_p$, corresponding to the portion of the disk provided as input. The parameter $R_p$, representing the putative planet’s radial location, is required as an additional input to rescale the observation to the training convention $(R_p=1)$, thereby ensuring consistency between the input image and the data used to train the model. No additional preprocessing or masking of secondary gaps is performed.

The more conservative approach would thus be to consider each result as an independent scenario where the disk has only one embedded planet at the specified location, producing all substructures within this disk region. However, although a systematic test is still pending, we can reasonably assume that the inferred posteriors are mainly informed by the closest substructures to the planet location, as was demonstrated on a similar CNN model by \cite{Zhang2022PGNets:Discs}. If this were true, our estimates would still be valid for a system with multiple planets, given that the effect of their mutual interaction on the disk morphology can be neglected. We only use this assumption in Sect. \ref{fig:stability} to speculate about the configuration of these potentially multiple systems. 
We also note that \citet{ShafaatMahmud2025InferringAutoencoder} developed an autoencoder-based pipeline (called VADER) to estimate planet masses from dust substructures in protoplanetary disks, trained on synthetic observations containing up to three planets. In their study, they compared the mean planet-mass predictions from VADER and DBNets \citep{Ruzza2024DBNets:Discs} for multi-planet and single-planet systems and found no statistically significant difference between the two groups.

\section{Details of sample selection and relevant literature}
\label{app:allsources}
In the following we discuss, for each source, our choices for the proposed planet locations reviewing the continuum radial profiles published in \cite{Curone2025ExoALMA.Emission} and relevant literature.

\subsection{AA Tau} 
This disk exhibits 3 pairs of gaps and rings in the \texttt{frank} radial profile, along with one fainter outer pair. Both \cite{Loomis2017ATau} and \cite{Curone2025ExoALMA.Emission} present evidence suggesting a misalignment of the inner disk. We propose D11 (11 au) as the first location for a putative planet. The two gap-ring pairs D64-B72 and D80-B90 correspond to only one gap in the CLEAN data. Notably, \cite{Pinte2025ExoALMA.Planetsb} proposed a kink detection at this location (B72), although with a low ($\sim 5$) signal-to-noise ratio. Other studies, such as \cite{Wang2021ArchitectureGap} have considered this substructure as a unique planet-hosting gap. Therefore, we assume 72 au, coincident with the ring in the \texttt{frank} profile, as our fiducial location for a second putative planet.
The outer gap-ring substructure is too faint for our analysis. 

\subsection{CQ Tau} This disk presents a cavity-like morphology and spiral-like nonaxisymmetric features consistent with the prominent spirals observed in SPHERE scattered-light images by \cite{Hammond2022ExternalTau}. An inner companion is proposed as responsible for inducing such spirals. \cite{Curone2025ExoALMA.Emission} also revealed a faint inner disk. \cite{UbeiraGabellini2019AALMA} modelled this disk with hydrodynamical and radiative transfer simulations finding that a planet with a minimum mass of 6 to 9 M$_J$, located at 20~au from the star, could qualitatively reproduce the observed dust morphology. We assume 20~au as our proposed location for a putative planet. 

\subsection{DM Tau} This disk presents an inner gap at 14~au that we propose as the first location for a putative planet.  \cite{Willson2016SparseDiscs} found a potential point-source at $\sim 6$~au through sparse aperture masking interferometry, consistent with a planet of roughly 10 M$_J$. Two pairs of gap-ring substructures further from the central star are also observed. We assume these to be due to a single planet at 72~au consistently with other studies \citep{Uyama2017TheObjects, Wang2021ArchitectureGap, Francis2022GapTau}. \cite{Flaherty2020MeasuringSgr} line-broadening measures yielded results compatible with high levels of turbulence. 

\subsection{HD 135344B} This disk exhibits a central cavity ending with a bright ring at 51 au which might be eccentric. A second ring with a prominent azimuthal asymmetry is located at 78~au. The asymmetry can be explained with a vortex that might alone explain the disk morphology \citep{Cazzoletti2018EvidenceSpirals}. Infrared observations reveal a prominent spiral structure (e.g. \citealt{Muto2012DiscoveryTheory, Garufi2013SmallPlanet, Stolker2016ShadowsImaging}).
We propose two putative planets. One inside the cavity at 28 au, consistent with the location of the point-like feature observed by \cite{Maio2025UnveilingVLT/ERIS}. The second planet is proposed inside the gap at 66~au. \cite{Veronesi2019Multi-wavelengthMass} numerical modelling of the spiral structure observed in scattered light suggests that it may be due to two embedded planets. Additionally, the observation of a change in the pitch angle of the IR spiral and a filament in the continuum data, first observed by \cite{Casassus2021ASAO206462} and confirmed by \cite{Curone2025ExoALMA.Emission}, further supports the presence of the outer companion.

\subsection{HD 143006} This is a well studied source also present in the DSHARP \citep{Andrews2018TheOverview} survey.
The continuum observations show two gaps located at 22 and 52~au which we assume as possible locations for embedded planets. Both \cite{Pinte2020NineGaps} and \cite{Pinte2025ExoALMA.Planetsb} claim the detection of a possible planetary signature in the disk kinematics approximately at the location of the inner gap. \cite{Ballabio2021HDDisc} with hydrodynamical simulations suggested a 3-10 M$_J$ planet located at 32~au within the
dust gap observed in scattered light.

\subsection{HD 34282} This disk presents a central cavity and several faint outer gap-ring pairs which are only clearly distinguishable in the \texttt{frank} profile. \cite{DeBoer2021Possible34282} observed a single arm spiral in IR, which would be better explained by a low mass planet inside its location. Outer gaps are too faint to apply DBNets2.0, therefore, we only assume the presence of one planet at 47~au corresponding to a shoulder in the depleted inner cavity, as revealed by the \texttt{frank} radial intensity profile. 

\subsection{J1604}
This source exhibits a large cavity depleted both in gas and dust. A bright ring is located at 82~au. Kinematical analysis provides evidence for a warp and reveal non-Keplerian features, including a spiral-like structure with a small ($6-14^{\circ}$) pitch angle \citep{Stadler2023ADisk, Pinte2025ExoALMA.Planetsb}. \cite{Stadler2023ADisk} found these features to be consistent with a planet located at 42~au, which we thus also assume as a tentative planet location. \cite{Yoshida2025ExoALMA.Wings} modeled this disk's surface density finding a drops by a factor of $\sim2–2.5$ at $\sim60$~au. Using \cite{Kanagawa2015FormationRotation} formula and the measured gap depth they estimated a putative planet mass between 0.1 and 2 M$_J$.

\subsection{J1615} This disk presents three pairs of faint gaps and rings two of which are too faint to sustain a DBNets2.0 analysis. We consider as a putative planet location the remaining gap at 83~au. \cite{Pinte2025ExoALMA.Planetsb} propose the detection of a planet kinematical signature located outside the region of continuum emission. 

\subsection{J1842} has an internal cavity with shadows that suggest the presence of an unseen inner ring. There are not proposed detections of planet signatures within this cavity that could inform us on a plausible planet location. DBNets2.0 results would be too dependent on the location choice. A gap at 63~au is revealed by the \texttt{frank} fit but is too faint to yield realiable results. \cite{Pinte2025ExoALMA.Planetsb} proposes a planet at $r_p=$105~au which is outside the region of continuum emission. We thus remove this disk from our sample of fiducial planets, but we still provide results for different planet locations in Appendix C.

\subsection{J1852} shows an internal cavity with a faint ring at 19~au which was previously proposed by \cite{Villenave2019SpatialRXJ1852.3-3700}. The location of maximum depletion within the cavity, i.e. 31~au, is proposed as a possile planet location.

\subsection{LkCa 15}. This disk presents one cavity and one gap. Several studies focused on this source exploring different scenarios for planetary companions. We consider two fiducial planet locations. The first would be at 42 au consistent with literature claims and likely responsible for the B68 ring. The second planet would be in the gap at 86 au. \cite{Long2022ALMADisk} claimed that the observed substructures, including asymmetric features, are best reproduced with more than one planet. \cite{Leemker2022GasCavities} observationally constrained the gas surface density across the cavity. They found evident gas depletion inside 10~au which suggested might be due to a Jovian planet, whereas they observed high gas density in the dust-depleted region between 10 and 68~au that was interpreted as a result of a chain of low mass planets. \cite{Gardner2025ExoALMA.Cavity} modelled both continuum and gas observations using hydrodynamical and radiative transfer simulations, enforcing \cite{Leemker2022GasCavities} hypothesis for the cavity and concluding that other plausible scenarios for the entire morphology could be either other processes that do not involve planets or three planets at 29, 51, and 80 au with masses of 3–5M$_J$, 0.4M$_J$, and 0.25M$_J$ respectively. \cite{Pinte2025ExoALMA.Planetsb} proposed detections of planet kinematic signatures which, however, lie outside of the continuum emission region. 

\subsection{MWC758} This disk presents a central cavity with a small inner disk and strong asymmetries that have been attributed to the global spiral structure clearly visible in scattered light observations \citep{Grady2013SpiralPlanets, Benisty2015Asymmetric758}. The spiral origin is disputed, but several studies argued for a planetary perturber in the outer disk \citep{Ren2018APlanets, Ren2020DynamicalDisk} with a tentative detection through high contrast imaging by \cite{Wagner2019ThermalArms}. However, \cite{Ren2018APlanets} and \cite{Calcino2020AreCavity} suggested that an eccentric perturber inside the cavity would also be possible. A point-like emission south of the star at a deprojected separation of 20~au was detected by \cite{Reggiani2018Discovery758} who suggested a CPD around a small mass companion between 0.5-5 M$_J$ as the more likely explanation. For this object, we use DBNets2.0 following the second hypothesis and therefore assuming a planet inside the cavity at 30~au at the location of the minimum in the \texttt{frank}-fitted radial profile emission. 

\subsection{SY Cha} This is a transition disk with dust morphology that resembles that of PDS70 while being less gas-depleted in the inner cavity/gap. The presence of an inner disk with an extension of a few au makes this a prototype disk for the formation of a solar system-like planetary system. Both \cite{Orihara2023ALMABand6SYChamaeleontis} and \cite{Curone2025ExoALMA.Emission} reported the observations of non-axisymmetric features in the dust continuum. We found no previous studies suggesting putative planets inside the gap. For our analysis, we assume a planet companion at a radial location of 33 au, corresponding to the gap minimum. 

\subsection{V4046 Sgr} This is a tight binary whose disk exhibits two concentric gaps with an inner disk that is off-centered with respect to the outer disk. This may be explained by a misalignment of the inner binary which could also account for the rest of the dust substructures \citep{Aly2020EfficientDiscs}. However, the planetary origin has also been explored. \cite{Ruiz-Rodriguez2019ConstraintsDisk} focused on the outer gap, checking its morphology against a set of hydrodynamical (and radiative transfer) simulations, concluding that the most likely explanation would be a Jovian planet. \cite{Weber2022TheV4046Sgr} also considered the inner gap successfully reproducing dust substructures in simulations with two embedded planets of 3.7 M$_J$ and 0.9~M$_J$ at 9 and 20~au respectively. Therefore, for our analysis we consider two fiducial planets at 8 and 20~au. However, caution is advised in this case as, if both planets were present, their small separation would more likely result in the substructures' morphology being affected by the presence of the second companion, a situation that is outside the scope of our tool.

\section{Degeneracy with the planet location within cavities}
\label{app:rpdeg}
\begin{figure}
    \centering
    \includegraphics[width=\linewidth]{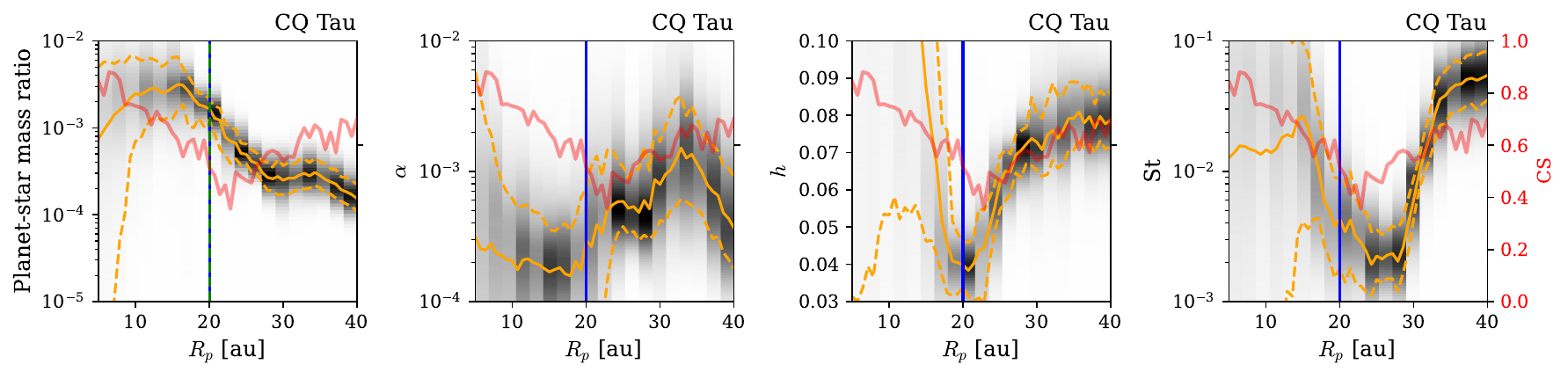}

    \includegraphics[width=\linewidth]{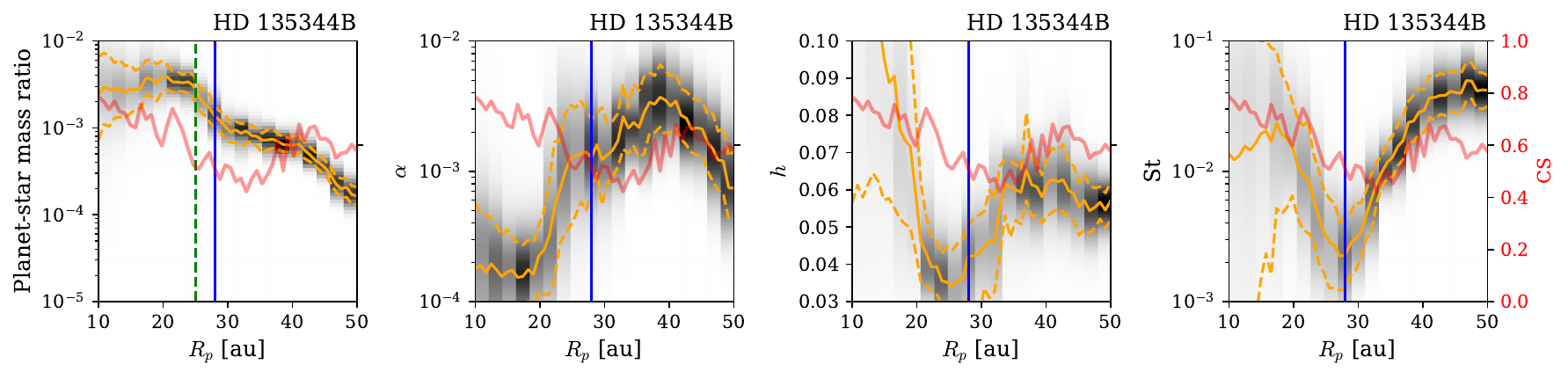}

    \includegraphics[width=\linewidth]{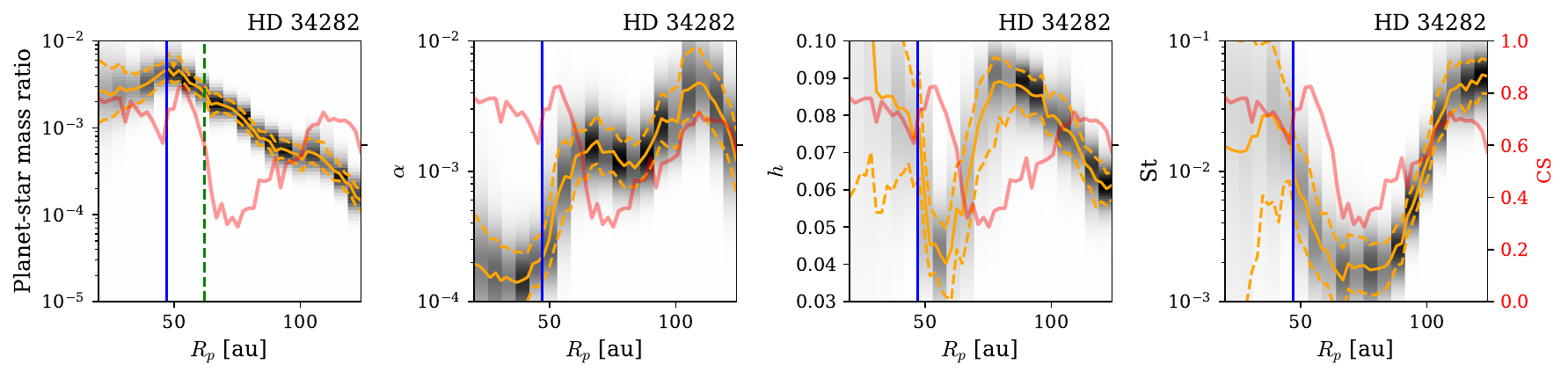}

    \includegraphics[width=\linewidth]{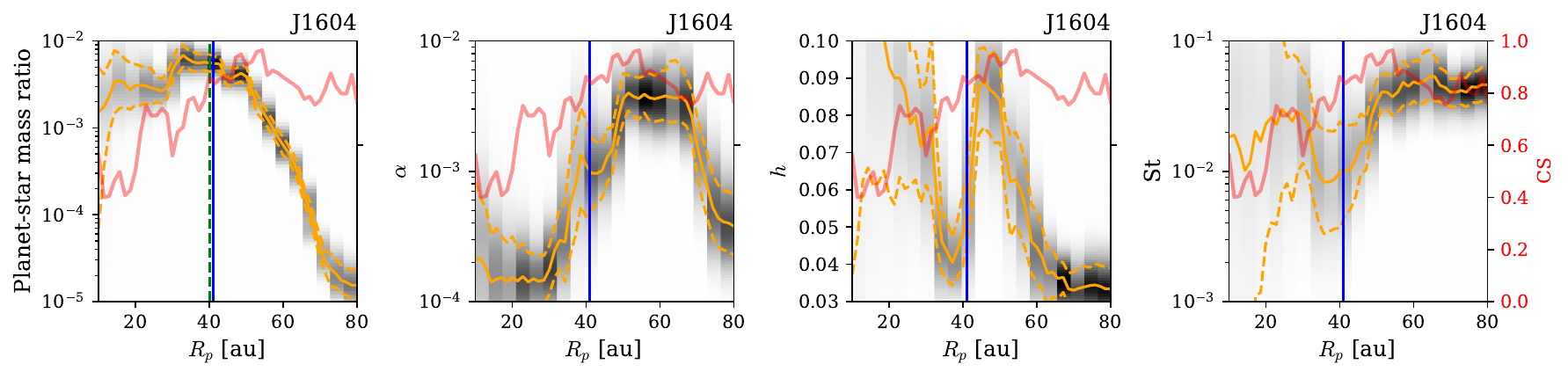}

     \includegraphics[width=\linewidth]{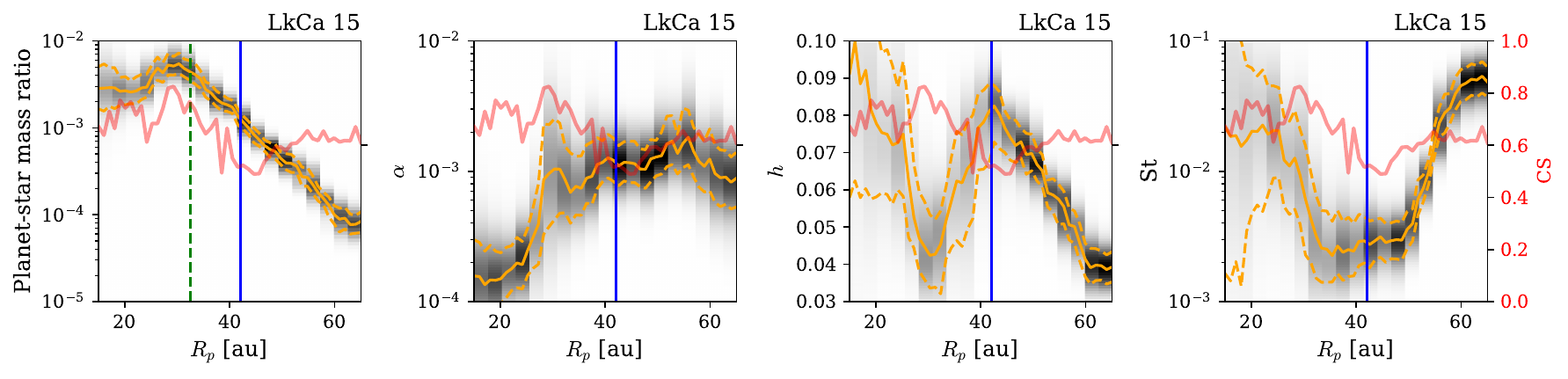}

    \caption{Sensitivity of DBNets2.0 inferred disk properties to the assumed planet location for all cavities considered in this study. The black 2D histograms represent the inferred distributions. The overlayed orange lines mark the 16th, 50th and 84th percentiles of the inferred posterior distributions. The vertical blue lines indicate the fiducial planet locations assumed in this work. The vertical green dashed lines mark $R_\text{edge}/2$, where $R_\text{edge}$ is the radial location of the cavity edge. The red lines indicate DBNets2.0 confidence score.}
    \label{fig:rpdeg}-
\end{figure}

\begin{figure}
    
    \centering
    \includegraphics[width=\linewidth]{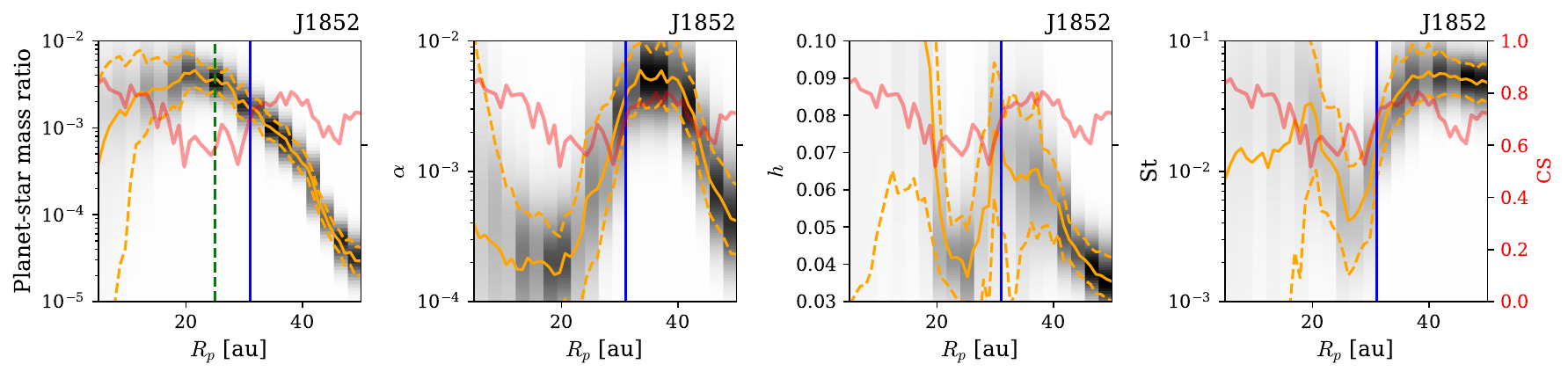}

    \includegraphics[width=\linewidth]{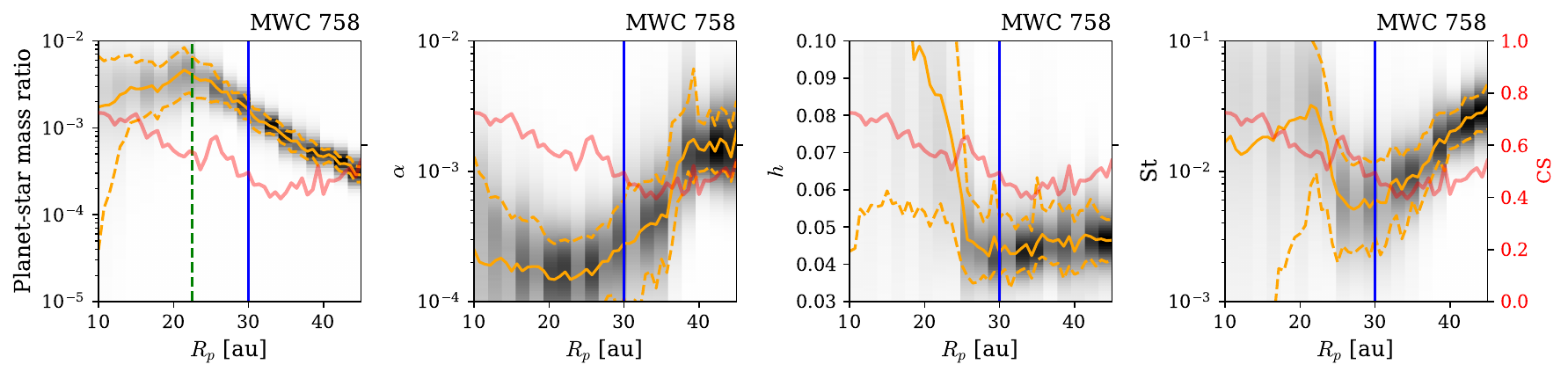}

    \includegraphics[width=\linewidth]{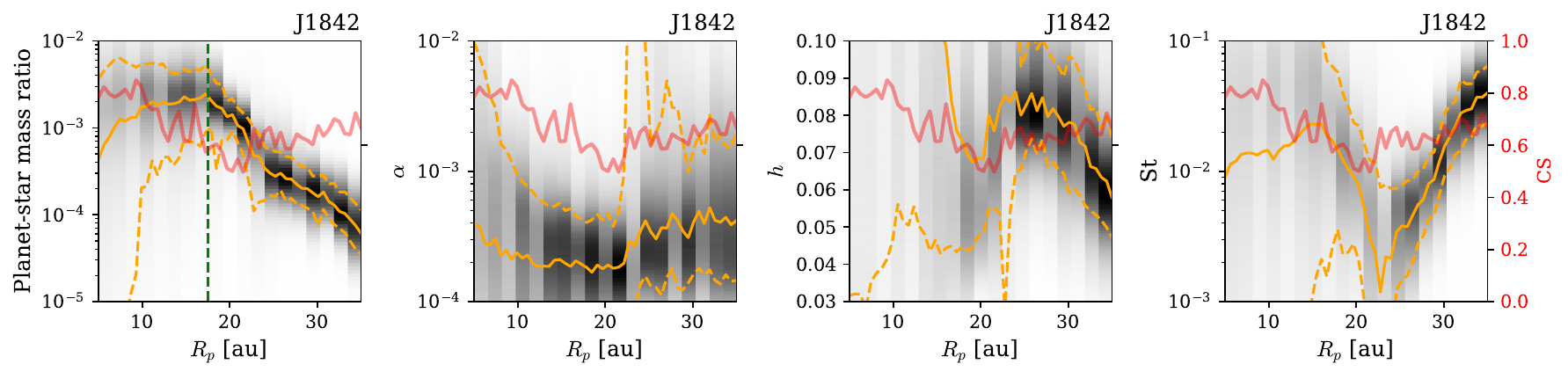}

    \caption{Sensitivity of DBNets2.0 inferred disk properties to the assumed planet location for the three peculiar cases of J1852, MWC~758, and J1842. In the two former disks, the considered substructures are large gaps with, respectively, a small inner disk and a faint inner ring. The latter disk presents a cavity, which was discarded from our analysis due to the lack of previous works suggesting possible planet locations. The black 2D histograms represent the inferred distributions. The overlayed orange lines mark the 16th, 50th and 84th percentiles of the inferred posterior distributions. The vertical blue lines indicate the putative planet location assumed in this work. The vertical green dashed lines mark $R_\text{edge}/2$, where $R_\text{edge}$ is the radial location of the cavity edge. The red lines indicate DBNets2.0 confidence score.}
    \label{fig:rpdeg2}
\end{figure}

We illustrate in Fig. \ref{fig:rpdeg} and \ref{fig:rpdeg2} how DBNets2.0 estimates for the disk properties $M_p, \alpha, h$, and $\text{St}$ vary as the planet is assumed in different positions inside the observed substructures. The variation of DBNets2.0 confidence score is also reported in the same figures. Figure \ref{fig:rpdeg} presents the results of this analysis for all disk's cavities where we assumed the presence of embedded planets. Figure \ref{fig:rpdeg2} presents three additional cases for which is worth performing this analysis, namely J1852, MWC758, and J1842. In the two former cases, putative planets are located within large gaps with an inner boundary defined, respectively, by a small inner disk and by a faint ring. These peculiar morphologies make the putative location of embedded planets uncertain.
The latter disk, instead, presents an inner cavity but was excluded from our sample due to the lack of other works or evidence suggesting a possible planet location. 

\section{All individual results}
\label{app:all_res}
\begin{figure}[p!]
    \centering
\includegraphics[width=0.49\linewidth]{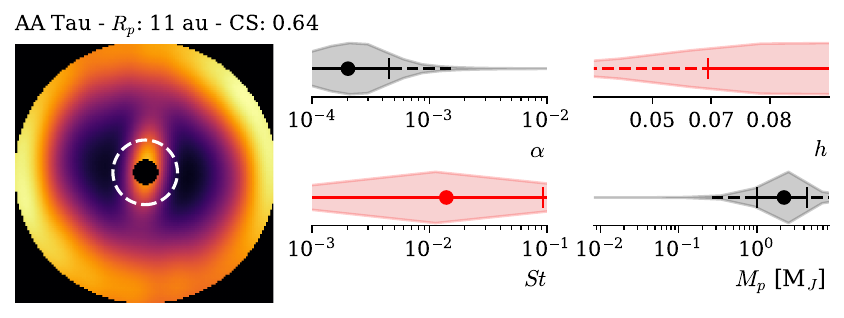}
\includegraphics[width=0.49\linewidth]{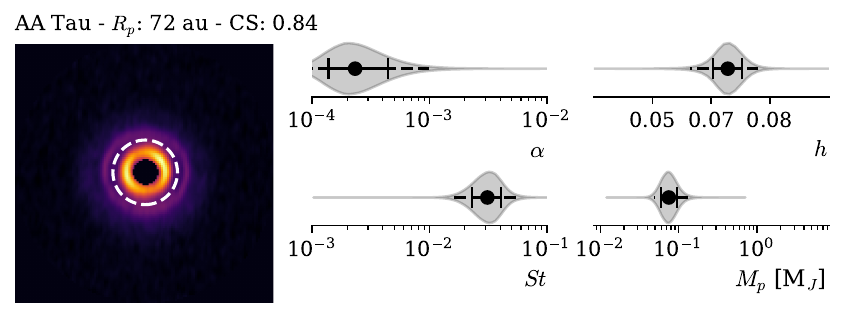}
\includegraphics[width=0.49\linewidth]{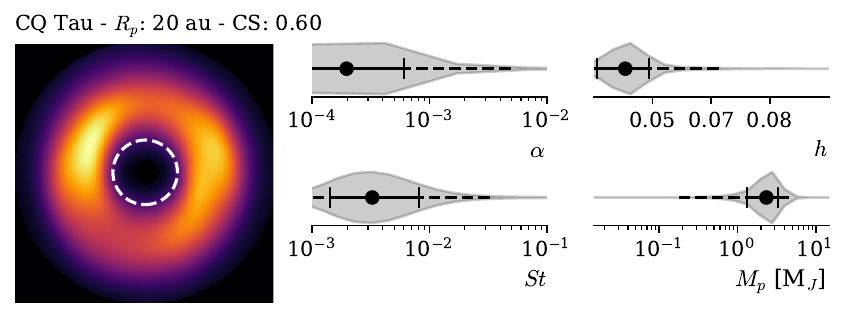}
\includegraphics[width=0.49\linewidth]{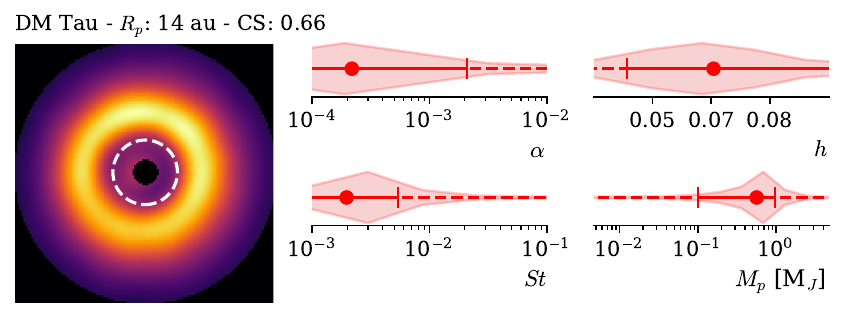}
\includegraphics[width=0.49\linewidth]{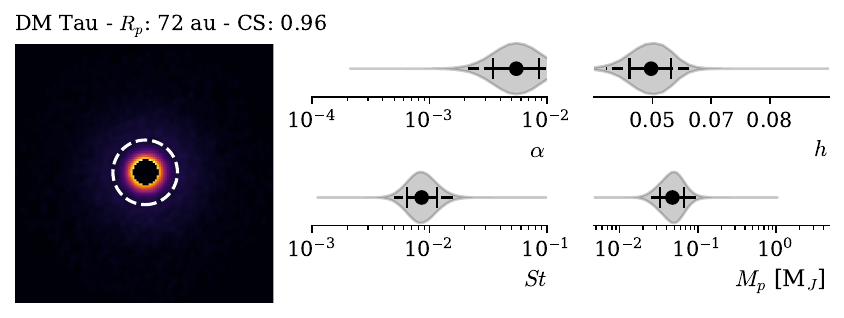}
\includegraphics[width=0.49\linewidth]{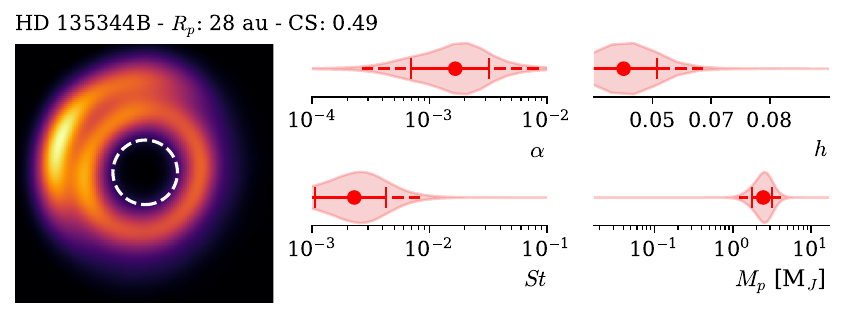}
\includegraphics[width=0.49\linewidth]{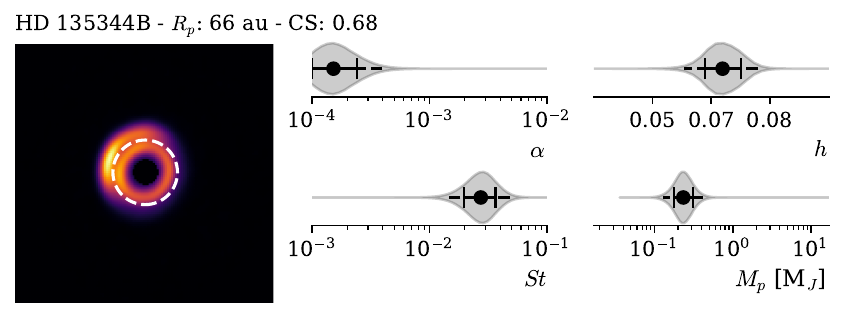}
\includegraphics[width=0.49\linewidth]{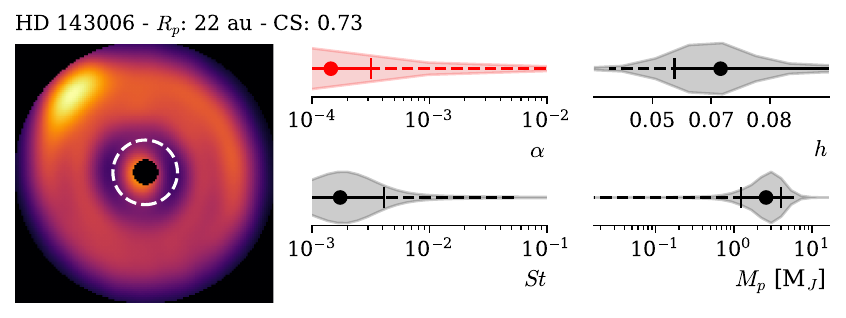}
\includegraphics[width=0.49\linewidth]{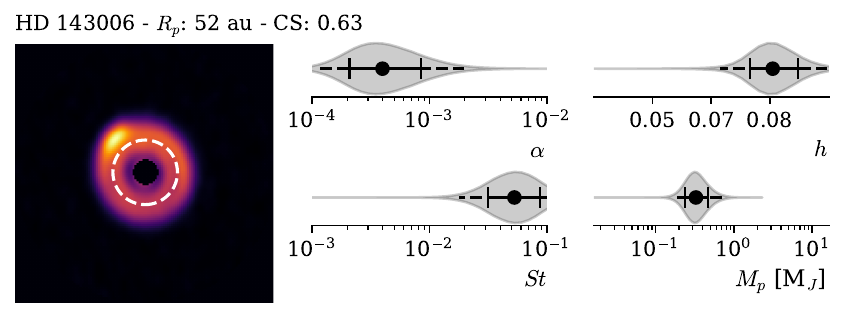}
\includegraphics[width=0.49\linewidth]{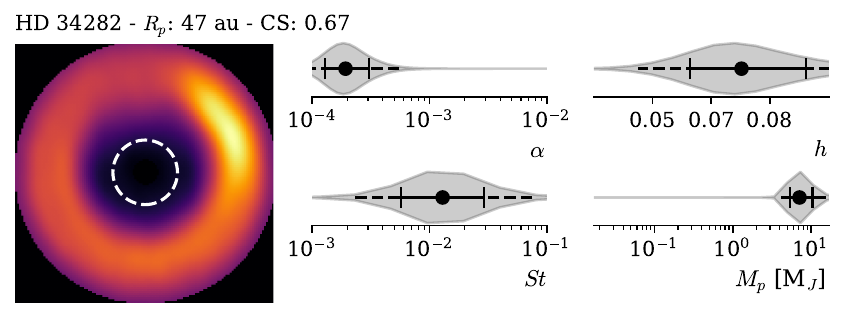}
    \caption{First gallery of DBNets2.0 results on the considered sample, continues in Fig. \ref{fig:allresb}. All disk images are shown after deprojection and rescaling. Disk names, fiducial planet positions ($R_p$, marked with the white dashed circle), and confidence scores (CS) are reported in the panels' titles. The violin plots represent DBNets2.0 inferred posterior distributions for each disk and planet property; the continuous error bars inside the violins mark the 16th, 50th, and 84th percentiles, while the dashed lines indicate the range between the 2.5th and 97.5th percentiles. Red violins indicate estimates that do not meet our acceptance criteria, namely CS $> 0.6$ and both the 2.5th and 97.5th percentiles falling within the uniform prior of the respective property.}
    \label{fig:allresa}
\end{figure}

\begin{figure}[p!]
    \centering
\includegraphics[width=0.49\linewidth]{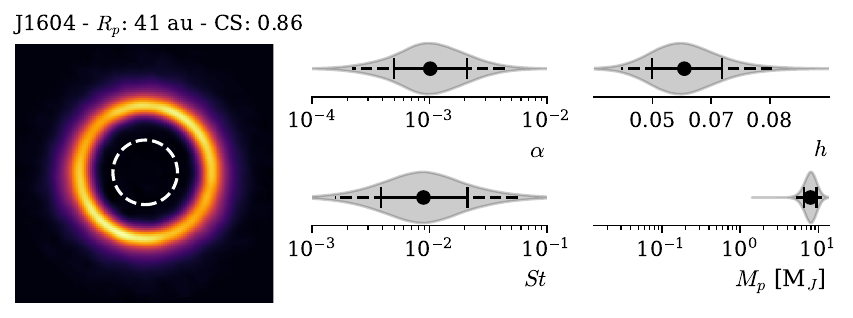}
\includegraphics[width=0.49\linewidth]{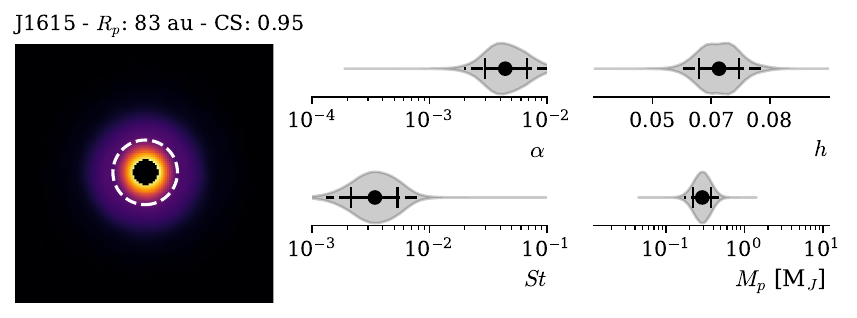}
\includegraphics[width=0.49\linewidth]{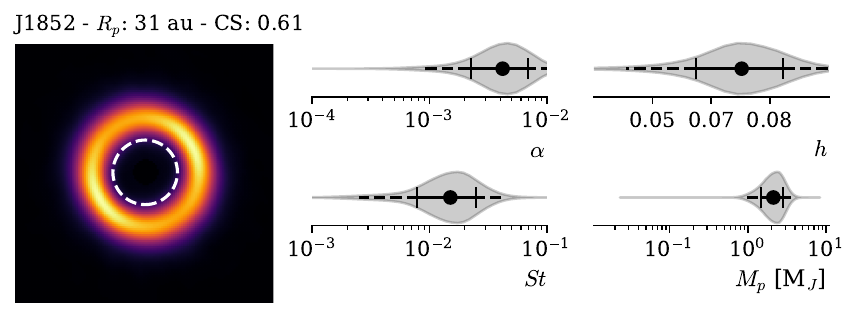}
\includegraphics[width=0.49\linewidth]{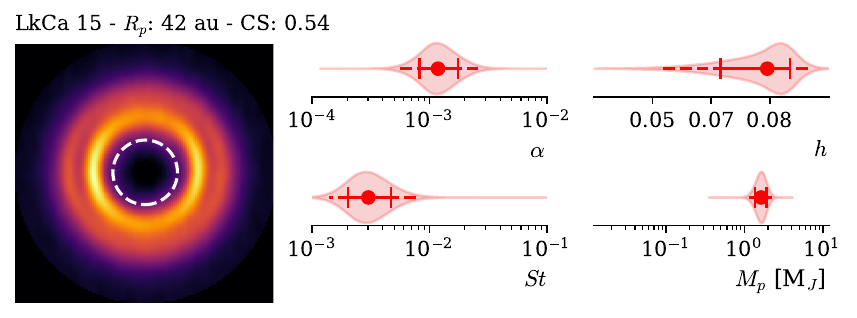}
\includegraphics[width=0.49\linewidth]{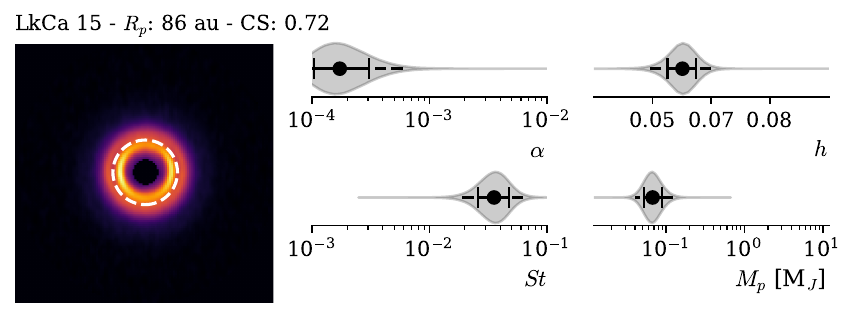}
\includegraphics[width=0.49\linewidth]{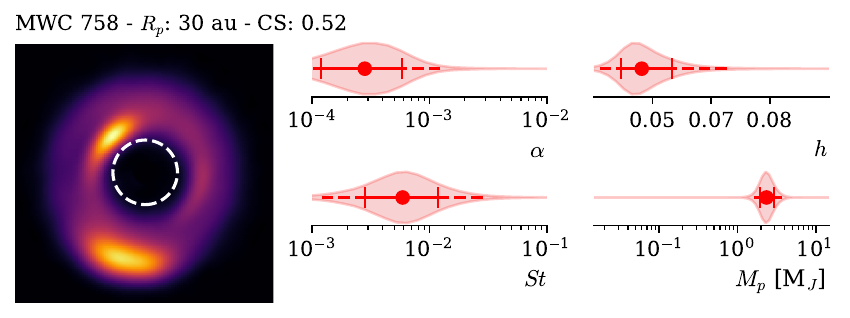}
\includegraphics[width=0.49\linewidth]{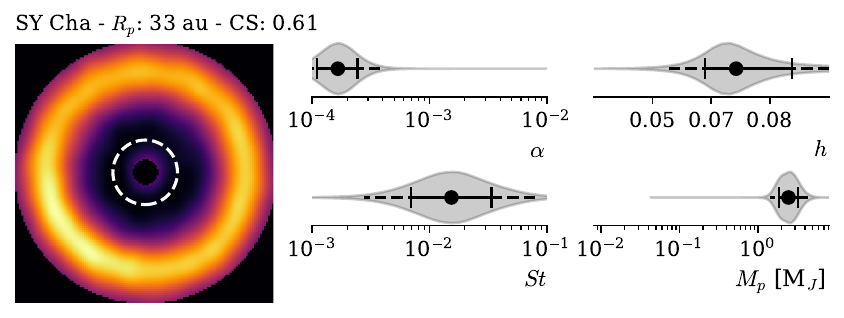}
\includegraphics[width=0.49\linewidth]{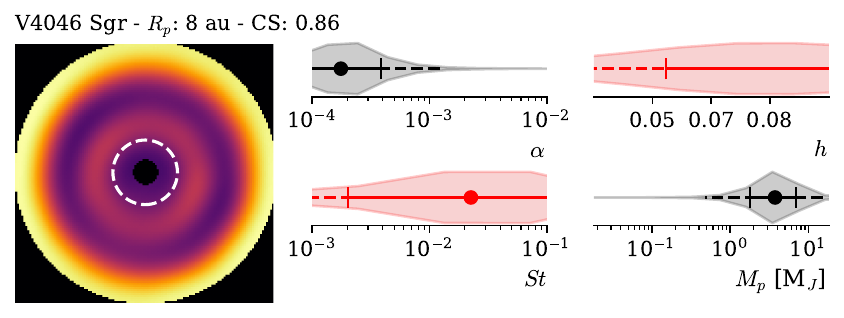}
\includegraphics[width=0.49\linewidth]{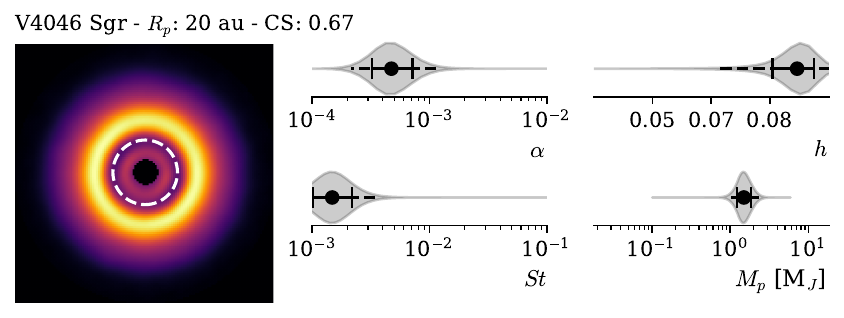}
\caption{Second gallery of DBNets2.0 results on the considered sample, continues from Fig. \ref{fig:allresa}. All disk images are shown after deprojection and rescaling. Disk names, fiducial planet positions ($R_p$, marked with the white dashed circle), and confidence scores (CS) are reported in the panels' titles. The violin plots represent DBNets2.0 inferred posterior distributions for each disk and planet property; the continuous error bars inside the violins mark the 16th, 50th, and 84th percentiles while the dashed lines indicate the range between the 2.5th and 97.5th percentiles. Red violins indicate estimates that do not meet our acceptance criteria, namely CS $> 0.6$ and both the 2.5th and 97.5th percentiles falling within the uniform prior of the respective property.}
    \label{fig:allresb}
\end{figure}

Figures \ref{fig:allresa} and \ref{fig:allresb} present singularly all the results of DBNets2.0 application on the selected targets. Best estimates with uncertainties are also listed in Table \ref{tab:allres}.

\section{Comparison with estimates in DBNets2.0 paper}
\label{app:compr25}
\begin{figure*}
\centering
    \includegraphics[width=\linewidth]{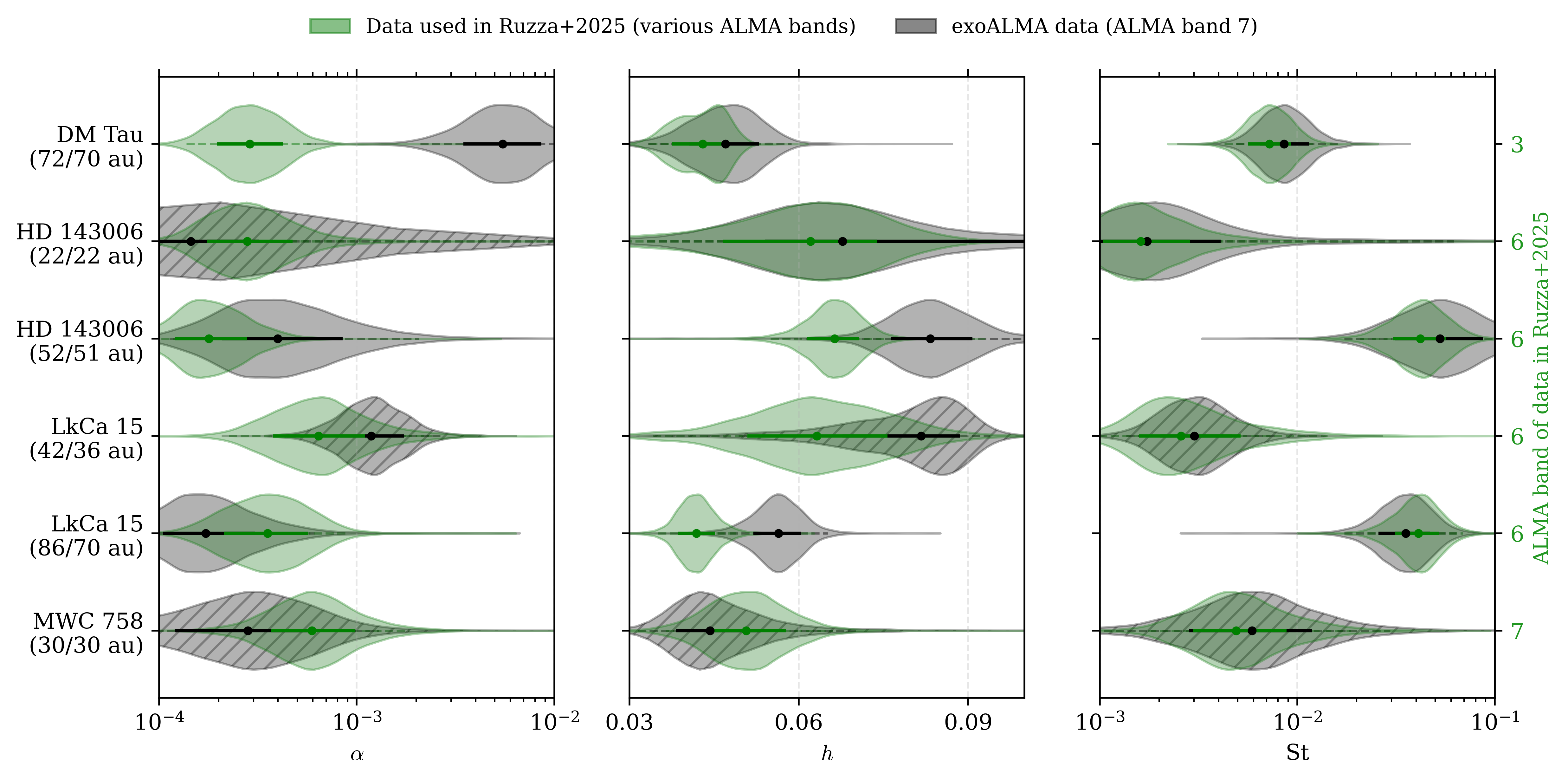}
    \caption{Comparison, for the overlapping objects, of DBNets2.0 estimates derived in \citetalias{Ruzza2025DBNets2.0:Discs} on archival data (green violins) with those obtained in this work on the exoALMA observations (black violins). On the right, in green, the Band of the continuum observation used in \citetalias{Ruzza2025DBNets2.0:Discs}. Near each disk name, we report the assumed planet locations, respectively, in \citetalias{Ruzza2025DBNets2.0:Discs} and in this work. Violin plots of unconstrained or unreliable estimates are hatched.}
    \label{fig:compR25}
\end{figure*}

We showed in Fig. \ref{fig:comp_ruzza25} the comparison between DBNets2.0 estimates for the planet mass obtained using exoALMA Band 7 observations with those obtained using better resolved, or shorter wavelength observations, in \citetalias{Ruzza2025DBNets2.0:Discs}. In Fig. \ref{fig:compR25}, we complete this comparison showing the other estimated properties, which we already discussed in Sect. \ref{sec:compR25}.

\section{Correlations with other disk properties}
\label{app:correlations}

\begin{figure}
    \centering
    \includegraphics[width=0.5\linewidth]{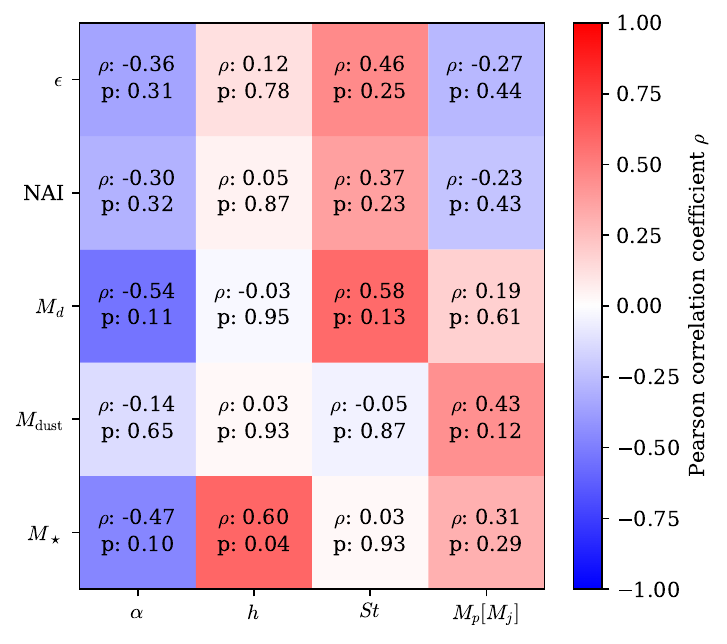}\hfill
    \includegraphics[width=0.4\linewidth]{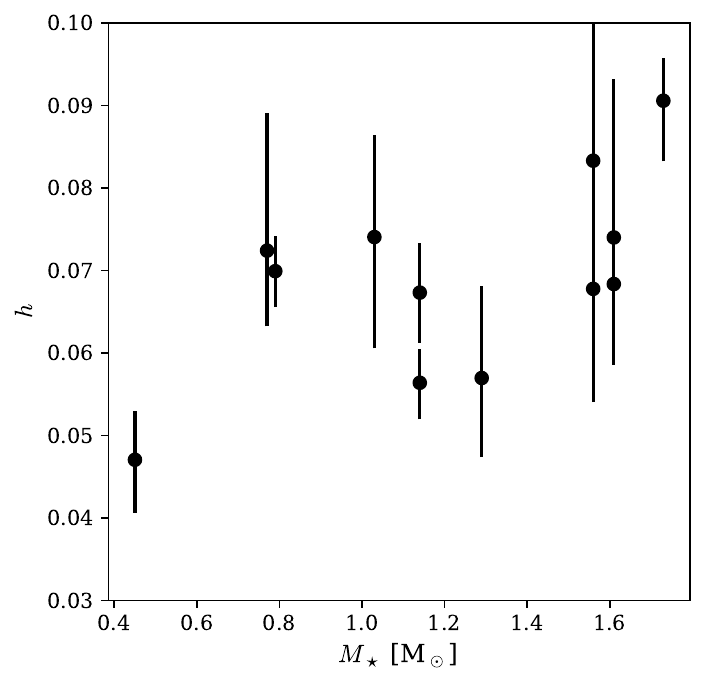}
    \caption{Left panel: Analysis of correlations between the estimated disk and planet properties with other disk features. Namely, from top to bottom, the disk gas-to-dust mass ratio ($\epsilon$) \citep{Curone2025ExoALMA.Emission}, the disks non-axisymmetry-index (NAI) \citep{Curone2025ExoALMA.Emission}, the disk dynamical masses \citep{Longarini2025ExoALMA.Modelling}, the disk dust masses \citep{Curone2025ExoALMA.Emission}, and the star masses \citep{Izquierdo2025ExoALMA.Disks}. The color bar refers to the Pearson correlation coefficient $\rho$. These are also reported in the plot for each combination of properties, together with the p-value corresponding to the null-hypothesis of the two properties being independent and normally distributed. Right panel: detail of the only statistically significant correlation highlighted by the analysis of Pearson correlation coefficients. This is between the inferred aspect ratios ($h$) and the host star masses.}
    \label{fig:correlations}
\end{figure}

We checked for correlations of the inferred disk and planet properties with other disk characteristics measured in separate studies, namely the disks’ gas-to-dust mass ratio, non-axisymmetry index, and masses of the gas, dust, and host stars. To do that, selected a pair of properties we computed the Pearson correlation coefficient ($\rho$), whose value indicates the strength and direction of a linear correlation (ranging from -1 to +1), and evaluated its statistical significance through the corresponding p-value. The latter indicates the probability of uncorrelated data to produce a Pearson correlation at least as extreme as the one computed. A p-value lower than 0.05 indicates statistical significant evidence for correlation between the two properties. For this analysis, we remove all estimates that we considered either unreliable or unconstrained. Figure \ref{fig:correlations} presents the results of this analysis. We find only one statistically significant correlation between stellar mass and inferred aspect ratio, which likely arises because more massive stars are more luminous, and thus their stronger irradiation heats the disk, increasing its aspect ratio.
\bibliography{references}{}

\bibliographystyle{aasjournalv7}



\end{document}